\documentclass[12pt]{emulateapj}

\shorttitle{Atomic Diffusion and Mixing in Old Stars I.}
\shortauthors{Korn et al.}

\begin{document}


\title{Atomic Diffusion and Mixing in Old Stars I.\\
VLT/FLAMES-UVES Observations of Stars in NGC 6397\altaffilmark{1}}


\author{A. J. Korn}
\affil{Department of Astronomy and Space Physics, Uppsala University, Box 515, 75120 Uppsala, Sweden}
\email{akorn@astro.uu.se}

\author{F. Grundahl}
\affil{Department of Physics and Astronomy, University of Aarhus,
Ny Munkegade, 8000 Aarhus C, Denmark}
\email{fgj@phys.au.dk}

\author{O. Richard}
\affil{GRAAL-UMR5024/IPM (CNRS), Universit\'{e} Montpellier II, Place E. Bataillon, 34095 Montpellier, France}
\email{Olivier.Richard@graal.univ-montp2.fr}

\author{L. Mashonkina}
\affil{Institute of Astronomy, Russian Academy of Science, Pyatnitskaya 48, 119017 Moscow, Russia}
\email{lima@inasan.ru}

\and

\author{P. S. Barklem, R. Collet\altaffilmark{2}, B. Gustafsson, N. Piskunov}
\affil{Department of Astronomy and Space Physics, Uppsala University, Box 515, 75120 Uppsala, Sweden}
\email{barklem, remo, bg, piskunov @astro.uu.se}

\altaffiltext{1}{Based on observations carried out at the European Southern Observatory (ESO), Paranal, Chile, under programme ID 075.D-0125(A).}

\altaffiltext{2}{present address: Max-Planck-Institut f\"{u}r Astrophysik, Karl-Schwarzschild-Str.~1, 85748 Garching, Germany}




\begin{abstract}
We present a homogeneous photometric and spectroscopic analysis of 18 stars along the evolutionary sequence of the metal-poor globular cluster NGC 6397 ([Fe/H]\,$\approx$\,$-$2), from the main-sequence turnoff point to red giants below the bump. The spectroscopic stellar parameters, in particular stellar-parameter differences between groups of stars, are in good agreement with broad-band and Str\"{o}mgren photometry calibrated on the infrared-flux method. The spectroscopic abundance analysis reveals, for the first time, systematic trends of iron abundance with evolutionary stage. Iron is found to be 31\,\% less abundant in the turnoff-point stars than in the red giants. An abundance difference in lithium is seen between the turnoff-point and warm subgiant stars. The impact of potential systematic errors on these abundance trends (stellar parameters, the hydrostatic and LTE approximations) is quantitatively evaluated and found not to alter our conclusions significantly. Trends for various elements (Li, Mg, Ca, Ti and Fe) are compared with stellar-structure models including the effects of atomic diffusion and radiative acceleration. Such models are found to describe the observed element-specific trends well, if extra (turbulent) mixing just below the convection zone is introduced. It is concluded that atomic diffusion and turbulent mixing are largely responsible for the sub-primordial stellar lithium abundances of warm halo stars. Other consequences of atomic diffusion in old metal-poor stars are also discussed.
\end{abstract}





\section{Introduction} \label{intr}
Since the pioneering work of \citet{Chamberlain_Aller_1951} stars have been used to trace the chemical evolution of the Milky Way and other galaxies. Due to their long lifetimes, cool stars (of spectral types F to K) are particularly suited for studying the build-up of elements over the course of the past 12-14\,Gyr. One of the basic assumptions made in this field of study is that the chemical composition in the atmospheric layers of such stars is representative of the gas out of which the stars once formed. In other words, it is assumed that there are no processes that substantially alter the chemical composition at the surface with time. However, stars are dynamical systems and various processes (e.g., convection and diffusion) can lead to an exchange of material between the star's interior and its atmosphere.

From the 1970ies, such processes have indeed been observed. For example, studies of carbon, nitrogen and oxygen abundances in giants stars \citep{Lambert_Ries_1977, Kjaergaard_etal_1982} show clear signs of mixing with material processed in the stellar core, in qualitative agreement with expectations from stellar structure and evolution \citep{Iben_1964}. In globular clusters, anti-correlations between certain elements (e.g.~O-Na) were also identified among giant stars \citep{Kraft_etal_1997}. It was, however, initially unclear whether these are caused by internal mixing processes or external pollution from an earlier generation of more massive stars. With the advent of 8m-class telescopes during the 1990ies, these anti-correlations could be traced in unevolved (turnoff-point) stars \citep{Gratton_etal_2001} clearly favouring the external-pollution scenario.

Beyond such element-specific effects, models of stellar evolution predict general variations of the atmospheric abundances of elements in unevolved stars on long timescales. Atomic diffusion is expected to be at work in the Sun \citep{Proffitt_Michaud_1991}, but larger effects are expected in warm metal-poor stars, as they are generally older and their convective envelopes are thinner. In contrast, giant stars are predicted not to show this effect, as their deep outer convection zones restore the original composition in their atmospheres. From this point of view, giant stars are to be preferred as probes of chemical evolution. However, their atmospheric structures may be more poorly known and certain (fragile) elements (e.g., the light elements Li, Be and B) can only be studied in near-main-sequence stars.

\begin{figure}[!t]
\includegraphics[angle=0,scale=.57]{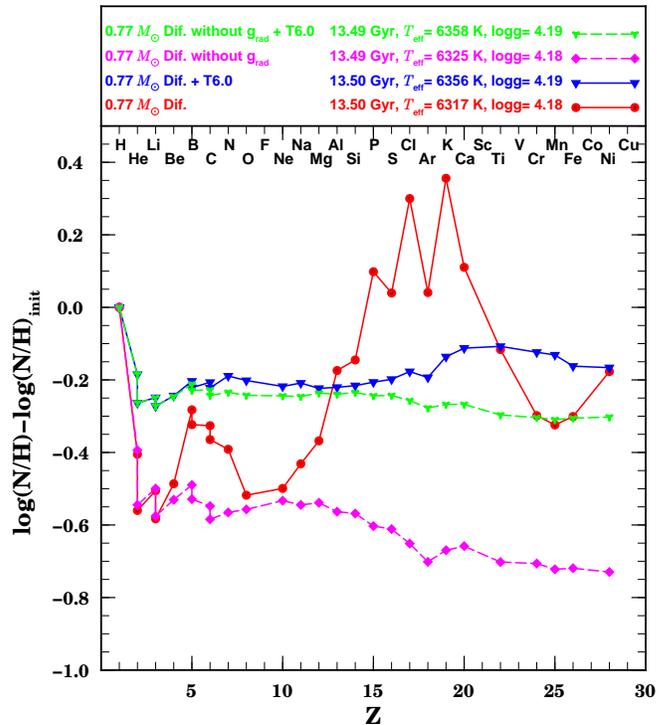}
\caption{Predictions for the variations of atmospheric abundances relative to the initial composition for a 0.77\,M$_\odot$ star with [Fe/H]\,=\,$-$2 that reaches the TOP after 13.5\,Gyr. Different symbols/line styles represent different models treating: atomic diffusion including radiative acceleration $g_{\rm rad}$ (bullets); atomic diffusion without radiative acceleration (rhombi); atomic diffusion including radiative acceleration and turbulent mixing T6.0 (triangles/full line); atomic diffusion without radiative acceleration, but with turbulent mixing T6.0 (triangles/dashed line). For details, see {\S}\ref{compdiff} and \cite{Richard_etal_2002}.}
\label{predictions}
\end{figure}

While early models predicted abundance reductions as large as a factor of ten \citep{Michaud_etal_1984}, current models treating atomic diffusion (including radiative acceleration) and mixing below the convection zone suggest abundance effects smaller than a factor of two (see Figure~\ref{predictions}). Effects of this size are hard to trace observationally. One needs to compare abundances in unevolved and evolved stars all drawn from the same stellar population. Globular clusters of the Galactic halo offer adequate laboratories to put observational constraints on this theoretical expectation.

\citet{King_etal_1998} used the large light-collecting area of the Keck telescope to investigate the chemical composition of stars at the turnoff point (hereafter TOP) of M92. They note ``possible evidence for [Fe/H] differences within M92'', but the design of their study (only observing TOP stars) and the moderate data quality (signal-to-noise (S/N) ratios between 10 and 40) precluded firm conclusions. Gratton et al.~(2001, hereafter G01) were the first to perform a joint analysis of stars at the TOP and the base of the red-giant branch (hereafter bRGB) in two nearby globular clusters, NGC\,6397 and NGC\,6752, using UVES on the VLT. While the main aim of their study was a detailed investigation of the anti-correlations discussed above, they found no indication of variations in iron abundances between the TOP and the bRGB stars.

Due to the fundamental importance of this result for cosmochemical studies (e.g., the interpretation of stellar lithium abundances in the framework of Big-Bang nucleosynthesis), we started to re-investigate the data of G01 in 2003. Data-reduction problems in the effective-temperature determination were uncovered which, once corrected for, leave room for a decrease of 20\% in iron abundance (0.1 dex in log(Fe)) due to atomic diffusion between these two groups of stars \citep{Korn_etal_2004}. This finding prompted us to obtain new data on NGC\,6397 in early 2005. Results concerning lithium were published in \citet{Korn_etal_2006}, here we give a full account of the analysis.

This article is organized in the following way: in {\S}2, the observations and data-reduction techniques are discussed. The spectroscopic and photometric stellar-parameter determinations and spectroscopic abundance analyses are presented in {\S}3. In {\S}4, the abundance trends are compared with predictions from stellar evolution models including diffusion. Finally, in {\S}{\S}5 and 6 the results are put into astrophysical context and are summarized.

\section{Observations}
\subsection{Photometry and Target Selection}
The photometric data on which our observations are based have been obtained with the Danish 1.54m telescope on La Silla, Chile. Str\"{o}mgren
$uvby$ data were obtained in May 1997 during a two-week observing run and additional $BVI$ data were obtained in 2005. For all these imaging data the DFOSC instrument was used.

All photometric reductions were carried out using the suite of programs developed by Peter Stetson \citep{Stetson_1987, Stetson_1990,Stetson_1994}.
The procedures for standard-star observations and photometric calibration of the $uvby$ photometry employed here are identical to those described in \cite{Grundahl_etal_1998, Grundahl_etal_1999, Grundahl_etal_2000, Grundahl_etal_2002} and the reader is referred to these papers for further details. The uncertainties in the photometric zero points for $vby$ are of the order 0\fm01, as found in previous studies.

For the $BVI$ data we did not observe standard stars since P.~Stetson has provided a library of secondary standards through the CADC WWW pages \footnote{http://cadcwww.hia.nrc.ca/cadcbin/wdb/astrocat/stetson/ query/} which are present in our field.
We therefore adopted the stars from his list for calibrating the broad-band photometry, and thus these stars act as local standards in each frame.

\begin{table}[!t]
\caption{Median and maximum number of observations in $uvby$. }
\label{uvby}
\begin{center}
\begin{tabular}{ccc}
\tableline\tableline
  Filter & Median & Max\\
\tableline
   $u$ & 10& 16 \\
   $v$ & 11& 17 \\
   $b$ & 12& 20 \\
   $y$ & 17& 27 \\
\tableline
\end{tabular}
\end{center}
\end{table}

The total field covered by the $uvby$ observations is roughly circular with
a radius of 9\arcmin, and was made from several overlapping fields. It is located roughly 5.5\,arcmin to the west of the cluster center, at (RA,DEC) = (17 40 04, $-$53 39 14) in the {\it UCAC2} \citep{Zacharias_etal_2004} system. The
median and maximum number of observations are given for each filter in
Table \ref{uvby}. For $BVI$ we obtained 3 exposures in each filter.

Our $uvby$ photometric data were also used to select the spectroscopic
targets and since observations are carried out with the fibre-fed multi-object spectrograph FLAMES accurate
astrometry was needed. For this we used the {\it UCAC2} catalog
 at Vizier\footnote{http://vizier.u-strasbg.fr/viz-bin/VizieR}. DFOSC is a focal reducer
which may have distortion that could affect the astrometric precision. We therefore calibrated each image separately, removed the distortion  and subsequently combined all the images to obtain the astrometry for each target. The astrometric precision is of the order 0\farcs10 for individual stars. The stars, their designations and J2000 positions are given in Table \ref{table:phot}.

\subsection{FLAMES+UVES Observations}
The FLAMES instrument \citep{Pasquini_etal_2002} has a combined mode which allows the simultaneous observation of up to eight objects with UVES and 130 objects with GIRAFFE-MEDUSA. Here we only report on the observations with FLAMES-UVES, the analysis of the GIRAFFE-MEDUSA data will be presented in a forthcoming paper (Lind et al., in preparation).

\begin{figure}[!t]
\includegraphics[angle=0,scale=.4]{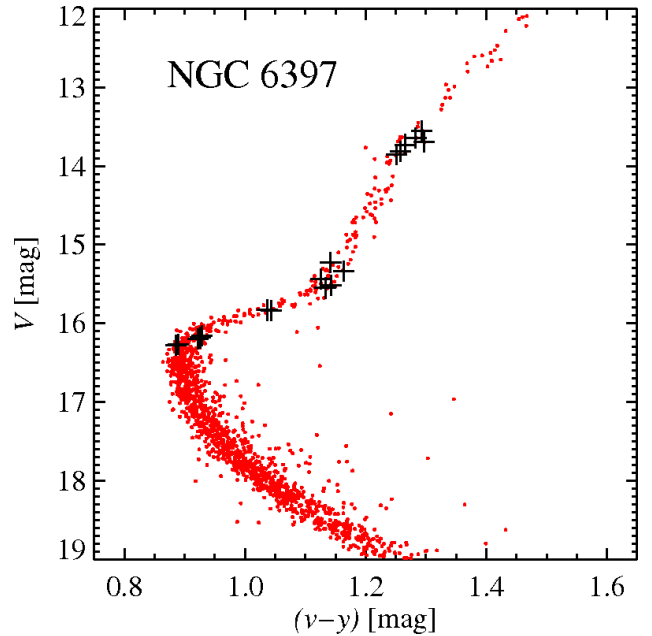}
\caption{Colour-magnitude diagram of NGC 6397 with the four groups of stars (from left to right TOP, SGB, bRGB and RGB stars) marked by crosses.}
\label{CMD}
\end{figure}

The total exposure time of $\approx$20.5\,h was split between three groups of stars, TOP, bRGB and RGB stars (see Figure \ref{CMD} and Table \ref{obs}). Two stars in the middle of the subgiant branch (hereafter SGB stars) were always observed in conjunction with the other groups of stars. With five to six stars in each group, this left one fibre to monitor the sky background. We re-observed some of the stars observed by G01, three in each of the two groups (TOP and bRGB).

The wavelength coverage of the FLAMES-UVES spectra is 4800\,--\,6800\,\AA, with a small CCD gap around 5800\,\AA. The resolving power is  $R$\,=\,47\,000. As seen in Table~\ref{obs}, the peak S/N per pixel varies between 56 (TOP star 507433) and 151 (RGB star 11093), while the average S/N in each group is more homogeneous and varies between 66 (TOP group) and 112 (SGB/RGB group). These values were computed from the rather line-free region around 5725\,\AA\ after rebinning by a factor of two (the rebinning does not degrade the resolution, as one resolution element consists of five pixels). The S/N remains roughly constant redward of 5725\,\AA, while
it gradually degrades (by $\leq$ 20\,\%) towards the blue end of the
spectrum at 4800\,\AA. In total, the S/N for the RGB stars is in line with the expectations from the FLAMES-UVES exposure-time calculator, while it falls short for the TOP stars by about 50\,\%. At least in part this can be attributed to the sky correction which inevitably adds noise.

For comparison, the five TOP stars observed with UVES in slit mode by G01 had a total integration time of 20\,h achieving S/N ratios between 60 and 90 at a resolving power of $R$\,=\,40\,000. With FLAMES-UVES, we thus profit from a significant multiplexing advantage (at the
dispense of wavelength coverage in the blue).

Barycentric radial velocities were computed from radial-velocity measurements using several strong Mg\,{\sc i} and Fe\,{\sc i} lines. As shown in Table~\ref{obs}, barycentric radial velocities of individual stars scatter around a mean value of 17.5\,km/s. This compares well with the mean radial velocity for NGC 6397 of 18.9\,km/s given by \citet{Harris_1996}. With typical velocity dispersions of 5-10\,km/s for old globular clusters, all stars are found to be kinematic cluster members. Our abundance analysis confirms cluster membership for all stars analysed.


\begin{table*}[!t]
\caption{Observational data of the NGC 6397 stars observed with FLAMES-UVES.}
\label{obs}
\begin{center}
\begin{tabular}{rrrcrrrcc}
\tableline\tableline
group & star id & star id & $V$ & $n_{\rm exp}$ & $t_{\rm exp}$ & S/N$_{\rm tot}$\tablenotemark{a} & corrected & $v_{\rm rad}^{\rm bary}$ $\pm$ 1$\sigma$\\
 & (this work) & (G01) & & & [h] & pixel$^{-1}$ & for sky & [km/s]\\
\tableline
TOP & 9655 & 1622 & 16.20 & 12 & 12.75 & 75 & yes & 16.1 $\pm$ 0.4 \\
TOP & 10197 & --- & 16.16 & 12 & 12.75 & 69 & yes & 13.4 $\pm$ 0.3 \\
TOP & 12318 & --- & 16.18 & 12 & 12.75 & 68 & yes & 19.1 $\pm$ 0.4 \\
TOP & 506120 & 202765 & 16.27 & 12 & 12.75 & 63 & yes & 24.4 $\pm$ 0.3 \\
TOP & 507433 & 201432 & 16.28 & 12 & 12.75 & 56 & yes & 20.7 $\pm$ 0.3 \\
SGB & 5281 & --- & 15.84 & 20 & 20.45 & 114 & yes & 18.8 $\pm$ 0.2\\
SGB & 8298 & --- & 15.83 & 20 & 20.45 & 110 & yes & 17.9 $\pm$ 0.2\\
bRGB & 3330 & --- & 15.22 & 6 & 6.20 & 97 & yes & 17.9 $\pm$ 0.1 \\
bRGB & 6391 & 793 & 15.55 & 6 & 6.20 & 75 & yes & 16.6 $\pm$ 0.1 \\
bRGB & 15105 & --- & 15.44 & 6 & 6.20 & 76 & yes & 19.6 $\pm$ 0.1 \\
bRGB & 23267 & 669 & 15.34 & 6 & 6.20 & 77 & yes & 17.1 $\pm$ 0.1 \\
bRGB & 500949 & 206810 & 15.51 & 6 & 6.20 & 76 & yes & 15.1 $\pm$ 0.1 \\
RGB & 4859 & --- & 13.82 & 2 & 1.50 & 117 & no & 17.9 $\pm$ 0.1 \\
RGB & 7189 & --- & 13.73 & 2 & 1.50 & 102 & no & 15.7 $\pm$ 0.1 \\
RGB & 11093 & --- & 13.55 & 2 & 1.50 & 151 & no & 19.1 $\pm$ 0.1 \\
RGB & 13092 & --- & 13.64 & 2 & 1.50 & 111 & no & 15.6 $\pm$ 0.1 \\
RGB & 14592 & --- & 13.70 & 2 & 1.50 & 104 & no & 15.6 $\pm$ 0.1 \\
RGB & 502074 & --- & 13.85 & 2 & 1.50 & 87 & no & 13.5 $\pm$ 0.1 \\
\tableline
\end{tabular}
\tablenotetext{a}{Computed from the $n_{\rm exp}$ co-added spectra in the wavelength range around 5725\,\AA}
\end{center}
\end{table*}

\subsection{Data Reduction}\label{reduction}
We make use of the FLAMES-UVES data-reduction pipeline. After careful inspection of the behaviour of the continuum, we found this pipeline to yield a sufficiently well-behaved run of the continuum to recover the intrinsic profile of lines as wide as or wider than the free spectral range (FSR) of UVES (70\,\AA\ at H$\alpha$). The fibre-feed to UVES improves the echelle-blaze removal, as both the star and the flat-field lamp illuminate the spectrograph in the same way. As shown in Fig.~\ref{Halpha}, this is not the case when UVES is used as a slit spectrograph: the blaze removal and subsequent order merging as implemented in the UVES pipeline leave significant residuals from which the analysis by G01 was shown to suffer \citep{Korn_etal_2004}.

The individual exposures are corrected for fibre-to-fibre throughput and the sky as monitored by one fibre is subtracted. They are co-added using MIDAS and C routines kindly made available by N.~Christlieb.

\section{Analysis} \label{ana}
We perform independent photometric and spectroscopic analyses to derive the fundamental stellar parameters $T_{\rm eff}$ and log $g$. While it is desirable to achieve stellar parameters as free of systematic errors as possible, the detection of abundance differences between groups of stars basically requires stellar-parameter {\em differences} that are free of systematic effects. Thus, for abundances derived from minority species like Li\,{\sc i}, Mg\,{\sc i} and Fe\,{\sc i}, $\Delta T_{\rm eff}$\,=\,$T_{\rm eff}$(TOP) $-$ $T_{\rm eff}$(RGB) is the most pertinent quantity, while majority species like Ti\,{\sc ii}, Fe\,{\sc ii} and Ba\,II react most sensitively to $\Delta \log g$ = log~$g$(TOP) $-$ log $g$(RGB).

\subsection{Photometry}
Apart from providing the spectroscopic targets for our observations, the photometry can also be used to estimate the basic stellar parameters,
such as effective temperature and surface gravity independently of the spectroscopic analysis. Here we describe the approach used to estimate these quantities.

It was found that our targets showed some scatter around the observed cluster sequences, most likely due to crowding effects and observational errors in the photometry. At the TOP, duplicity may play a role. In order to reduce this effect, we derived mean cluster fiducial sequences by first cleaning the photometry lists based on photometric quality (the DAOPHOT {\it SHARP\/} parameter) and location in the cluster (crowding). We then divided the $V$ magnitudes into 0\fm22 bins and for each of these bins the {\it same} stars were averaged (using robust estimation routines from the Astrolib library) for each filter ($uvbyBVI$) to produce mean fiducial points
for each band.

When deriving effective temperatures based on photometry we use colour indices, and to ensure that these properly represent the cluster sequence we overplotted the fiducial points on the cluster colour-magnitude diagram (CMD) for each colour-magnitude combination employed, for example a $(v-y,V)$ diagram. We estimate that the error in the colour difference between the RGB and the cluster TOP is less than 0\fm01 and we adopt this as our uncertainty estimate of this quantity.

Armed with the fiducial sequences we proceed to derive fiducial colours for our target stars by interpolating their colour
in the fiducial sequence at their observed $V$ magnitude. This approach effectively removes the star-to-star spread at any given magnitude and is the best we can do since our $V$ magnitudes are the best calibrated and have the highest number of observations per star.

\subsubsection{Effective temperatures}

To estimate the effective temperatures we use the empirical colour calibrations
by Alonso et al.\ (1996, 1999) and \cite{Ramirez_Melendez_2005}.
When using the $uvby$ system it has been found that the ($v-y$) index gives
a superior temperature discrimination for RGB and TOP stars compared to
$(b-y)$ which is normally used. Since such calibrations
had not been derived by either Alonso et al.\ or Ram\'{i}rez \& Mel\'{e}ndez we asked them to
derive new calibrations for different colour combinations and then used these (A.A. and I.R., priv. comm. 2006).

We present the new calibrations in Table \ref{photcalib} in the form of coefficients for the analytic formulae given in the respective original papers by Alonso et al.\ amd Ram\'{i}rez \& Mel\'{e}ndez.

Before applying the colour-temperature calibrations we discuss the cluster reddening since the calibrations employ the reddening-corrected colours. NGC 6397 has a non-negligible reddening and several estimates are available in the literature \citep{Harris_1996}. Following \citet{Anthony-Twarog_Twarog_2000}, we adopt
a value of E($B-V$)$\,=\,0.179\,\pm\,0.003$, a value very similar to
the one derived by G01. To correct for reddening for the other photometric bands employed here, we use the coefficients given in Table 1 of Ramirez \& Melendez (2005).

For each target star we then derived effective temperatures based on the available colours. We calculated each star's effective temperature using both the dwarf- and giant-star calibrations.  The resulting effective temperatures for each target are given in Table \ref{table:phot}.

For the calibrations used we find that the $(v-y)$ calibration has a sensitivity of 22\,K and 38\,K per 0\fm01 change in $(v-y)$ for
giants and turnoff stars at [Fe/H]$=\,-2.0$, respectively . The $(V-I)$ colour shows approximately the same sensitivity, i.e., with changes of around 20\,K and 40\,K for the giant and dwarf calibrations, respectively. For a change of $\pm$0.1\,dex in [Fe/H] the effective temperature changes by $\pm$20K. We have used a metallicity of $-$2.0 throughout for calculating effective temperatures, thus not making any implicit assumption about the existence of atomic diffusion.

As presented in Table \ref{table:phot}, the $\Delta T_{\rm eff}$ values vary between 906\,K ($B-V$) and 1108\,K ($b-y$). The other two calibrations both point to a $\Delta T_{\rm eff}$ close to 1100\,K, indicating that $B-V$ may underestimate the effective-temperature differences between TOP and RGB stars.

Larger $\Delta T_{\rm eff}$ values seem possible based on 2MASS
  $V-J$, $V-H$ and $V-K$ colours. However, at the magnitudes of TOP stars in
  NGC 6397, the 2MASS magnitudes have substantial uncertainties. More
  accurate infrared photometry would be desirable to reach the
  accuracy needed for studies like these.

\subsubsection{Surface gravities}

Surface gravities were derived from the usual relation between log $g$ and the luminosity, temperature and mass of the observed stars. We assumed a mass of 0.779\,M$_\odot$ for the TOP stars and 0.792\,M$_\odot$ for the RGB stars based on 13.5\,Gyr isochrones of \citet{Richard_etal_2005}. This is an upper limit for the age of NGC 6397 and the masses are therefore lower limits. The bias on the surface gravities introduced by these assumptions is, however, insignificant.

In order to determine the luminosity of the stars, we have to derive their bolometric corrections and
this was done using the calibrations of Alonso et al.~(1999) assuming a metallicity of $-$2.0 (as was done for the temperature derivation).
For our estimated uncertainties in $V$ (0\fm01), metallicity (0.1\,dex) and stellar mass (0.01\,M$_{\odot}$), the error
in the determination of the effective temperature is the dominant error source in the determination of the log $g$ {\it differences}. An error of 50\,K for the effective-temperature difference translates into a change of 0.015\,dex in $\Delta \log g$. For the calculation of the absolute surface gravities a value of $(m-M)_V\,=\,12.57$ was assumed. Given the reddening of the cluster, this value is well within the range determined by \cite{Reid_Gizis_1998}. We note that the assumed values of $(m-M)$ and reddening are secondary for the results of this work.

$\Delta \log g_{\rm phot}$ is determined to be 1.38 $\pm$ 0.05. The error is estimated from an uncertainty in effective temperature for each group of stars of 100\,K. Given the multiple observations and targets, this is a rather conservative error estimate.

\subsection{Spectroscopy} \label{spec}
\begin{figure}
\includegraphics[angle=90,scale=.35]{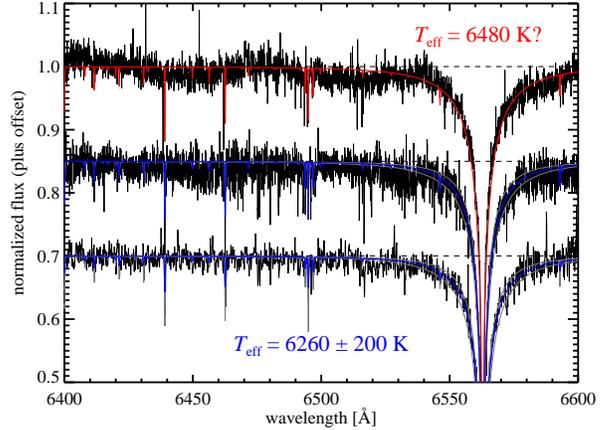}
\caption{Three spectra of the TOP star 507433 taken with UVES in slit mode ({\em top\/}), FLAMES-UVES ({\em middle\/}) and GIRAFFE-MEDUSA ({\em bottom\/}). All spectra are normalized using linear rectification curves. The UVES spectrum suffers from data-reduction problems as discussed in Sect.~\ref{reduction}. Both the FLAMES-UVES and the GIRAFFE-MEDUSA spectra indicate an effective temperature of $T_{\rm eff}$\,=\,6260\,K.}\label{Halpha}
\end{figure}
\begin{figure}
\includegraphics[angle=90,scale=.35]{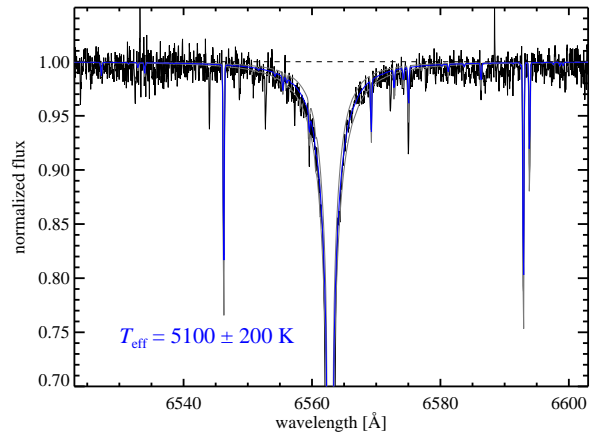}
\caption{FLAMES-UVES spectrum of the RGB star 11093. The Balmer line is significantly weaker than in the TOP stars, but the effective-temperature sensitivity coupled with the higher S/N ratio allows constraining $T_{\rm eff}$ to $\pm$200\,K.}\label{HalphaRGB}
\end{figure}

The spectroscopic analysis combines the constraints from fitting the wings of H$\alpha$ and the iron ionization equilibrium (Fe\,{\sc i/ii}) in non-LTE to iteratively derive reddening-free estimates of $T_{\rm eff}$, log $g$, the metallicity [Fe/H] and the microturbulence $\xi$. The analysis is line-by-line differential to the Sun and follows the techniques described in \citet{Korn_etal_2003}. It employs profile fits rather than equivalent widths.

The model-atmosphere code used is MAFAGS \citep{Grupp_2004} and employs the opacity-distribution functions (ODF) of \citet{Kurucz_1993}. To fulfil the constraints of Balmer line-profile fitting, a low value for the mixing-length parameter ($\alpha_{\rm MLT}$\,=\,0.5) is used \citep{Fuhrmann_etal_1993}. This choice has a minor effect on metal lines, as they originate from shallower optical depth. Of all the Balmer lines, H$\alpha$ is least affected by the choice of $\alpha_{\rm MLT}$\,=\,0.5. An enhancement of the $\alpha$-elements O, Mg and Si by 0.5, 0.4 and 0.4\,dex is used in the construction of the model atmosphere, as these elements can make significant contributions to the electron pressure and thus to the continuous opacity produced by H$^-$.

Test calculations with MARCS model atmospheres \citep{Gustafsson_etal_1975,Asplund_etal_1997} show very similar results, both for the effective temperatures (see Sect.~\ref{effteff}) and the relative abundances. This is not surprising, as the input physics is very similar in both programme suites. Differences in the opacity treatment (ODF vs.~opacity sampling) are not large at these low metallicities.

\begin{table*}[!t]
\caption{Spectroscopic stellar parameters of the 18 stars observed with FLAMES-UVES. The standard deviation for all Fe\,{\sc i} and Fe\,{\sc ii} lines measured (total number of lines in parentheses) is given. For $T_{\rm eff}$, log $g$ and $\xi$, measurement uncertainties of 200\,K, 0.2\,dex and 0.2\,km/s are assumed, respectively. These errors are propagated into the error for the group-averaged properties given at the bottom of the table.}
\label{stellparam}
\begin{center}
\begin{tabular}{rrrcccc}
\tableline\tableline
group & star id & star id & $T_{\rm eff}$ & log $g$ & [Fe/H] $\pm$ $1\sigma$ & $\xi$ \\
 & (this work) & (G01) & [K] & [cgs] & [dex] & [km/s] \\
\tableline
TOP & 9655 & 1622 & 6260 & 3.85 & $-$2.29 $\pm$ 0.06 (22) & 2.1 \\
TOP & 10197 & --- & 6250 & 3.95 & $-$2.29 $\pm$ 0.10 (21) & 2.0 \\
TOP & 12318 & --- & 6240 & 3.90 & $-$2.28 $\pm$ 0.08 (21) & 2.0 \\
TOP & 506120 & 202765 & 6260 & 3.85 & $-$2.26 $\pm$ 0.09 (23) & 1.9 \\
TOP & 507433 & 201432 & 6260 & 3.90 & $-$2.26 $\pm$ 0.06 (20) & 2.0 \\
SGB & 5281 & --- & 5800 & 3.55 & $-$2.25 $\pm$ 0.08 (26) & 1.75 \\
SGB & 8298 & --- & 5810 & 3.60 & $-$2.23 $\pm$ 0.07 (24) & 1.75 \\
bRGB & 3330 & --- & 5430 & 3.35 & $-$2.16 $\pm$ 0.08 (24) & 1.7 \\
bRGB & 6391 & 793 & 5510 & 3.40 & $-$2.14 $\pm$ 0.09 (29) & 1.6 \\
bRGB & 15105 & --- & 5470 & 3.40 & $-$2.18 $\pm$ 0.07 (25) & 1.75 \\
bRGB & 23267 & 669 & 5370 & 3.30 & $-$2.23 $\pm$ 0.09 (27) & 1.8 \\
bRGB & 500949 & 206810 & 5500 & 3.40 & $-$2.19 $\pm$ 0.09 (25) & 1.8 \\
RGB & 4859 & --- & 5150 & 2.65 & $-$2.12 $\pm$ 0.08 (38) & 1.6 \\
RGB & 7189 & --- & 5130 & 2.55 & $-$2.14 $\pm$ 0.08 (36) & 1.6\\
RGB & 11093 & --- & 5100 & 2.50 & $-$2.14 $\pm$ 0.08 (42) & 1.7\\
RGB & 13092 & --- & 5120 & 2.55 & $-$2.12 $\pm$ 0.08 (38) & 1.6\\
RGB & 14592 & --- & 5130 & 2.55 & $-$2.12 $\pm$ 0.08 (37) & 1.6\\
RGB & 502074 & --- & 5150 & 2.60 & $-$2.10 $\pm$ 0.09 (35) & 1.5\\
\tableline
TOP$_{\rm ave}$ & & & 6254 $\pm$ 90 & 3.89 $\pm$ 0.09 & $-$2.28 $\pm$ 0.04 & 2.0 $\pm$ 0.1\\
SGB$_{\rm ave}$ & & & 5805 $\pm$ 140 & 3.58 $\pm$ 0.14 & $-$2.24 $\pm$ 0.05 & 1.75 $\pm$ 0.15\\
bRGB$_{\rm ave}$ & & & 5456 $\pm$ 90 & 3.37 $\pm$ 0.09 & $-$2.18 $\pm$ 0.04 & 1.73 $\pm$ 0.1\\
RGB$_{\rm ave}$ & & & 5130 $\pm$ 80 & 2.56 $\pm$ 0.08 & $-$2.12 $\pm$ 0.03 & 1.6 $\pm$ 0.1\\
\tableline
\end{tabular}
\end{center}
\end{table*}

\subsubsection{Effective temperatures}\label{effteff}
\begin{figure}[!t]
\includegraphics[angle=90,scale=.35]{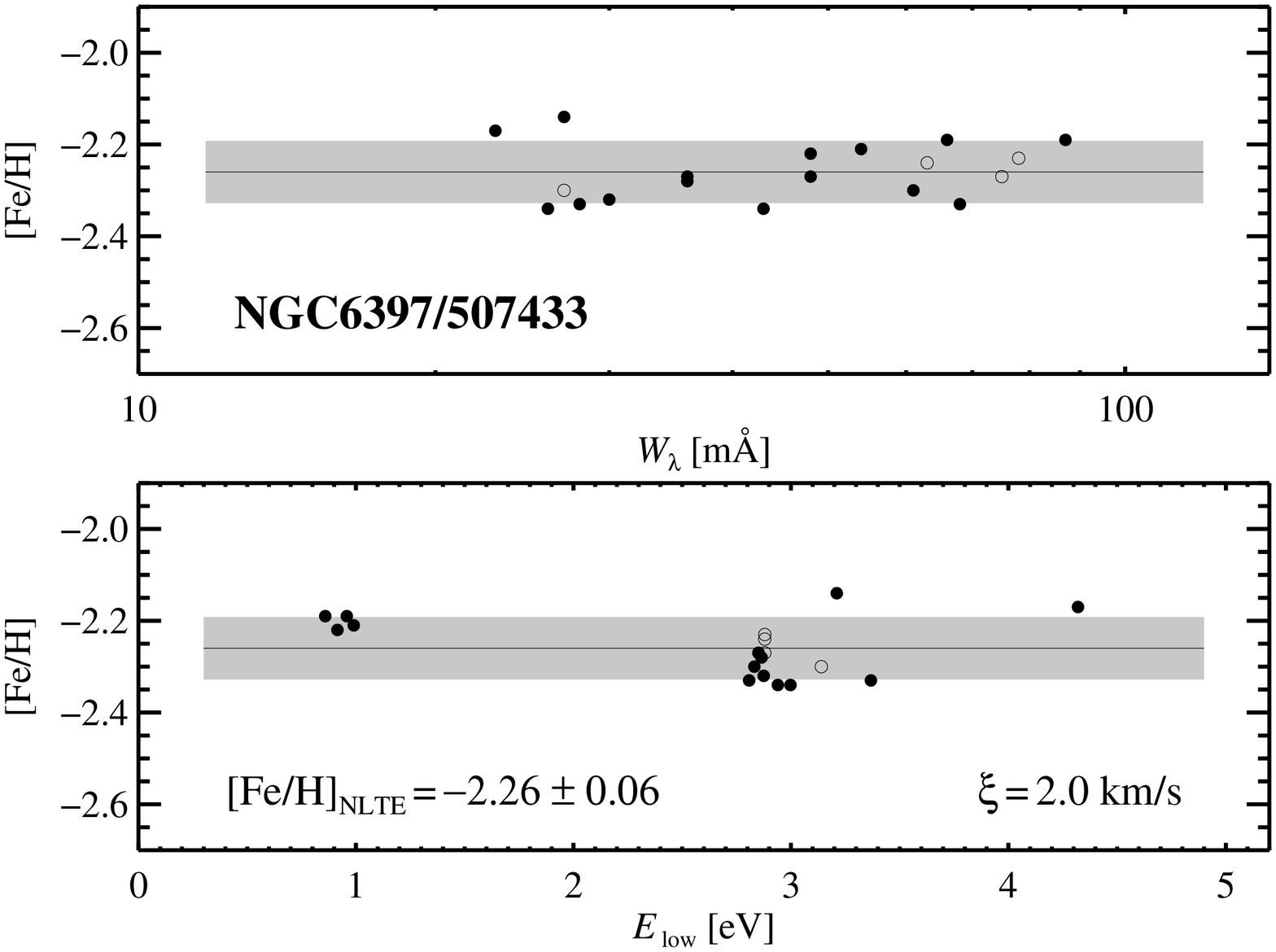}
\includegraphics[angle=90,scale=.35]{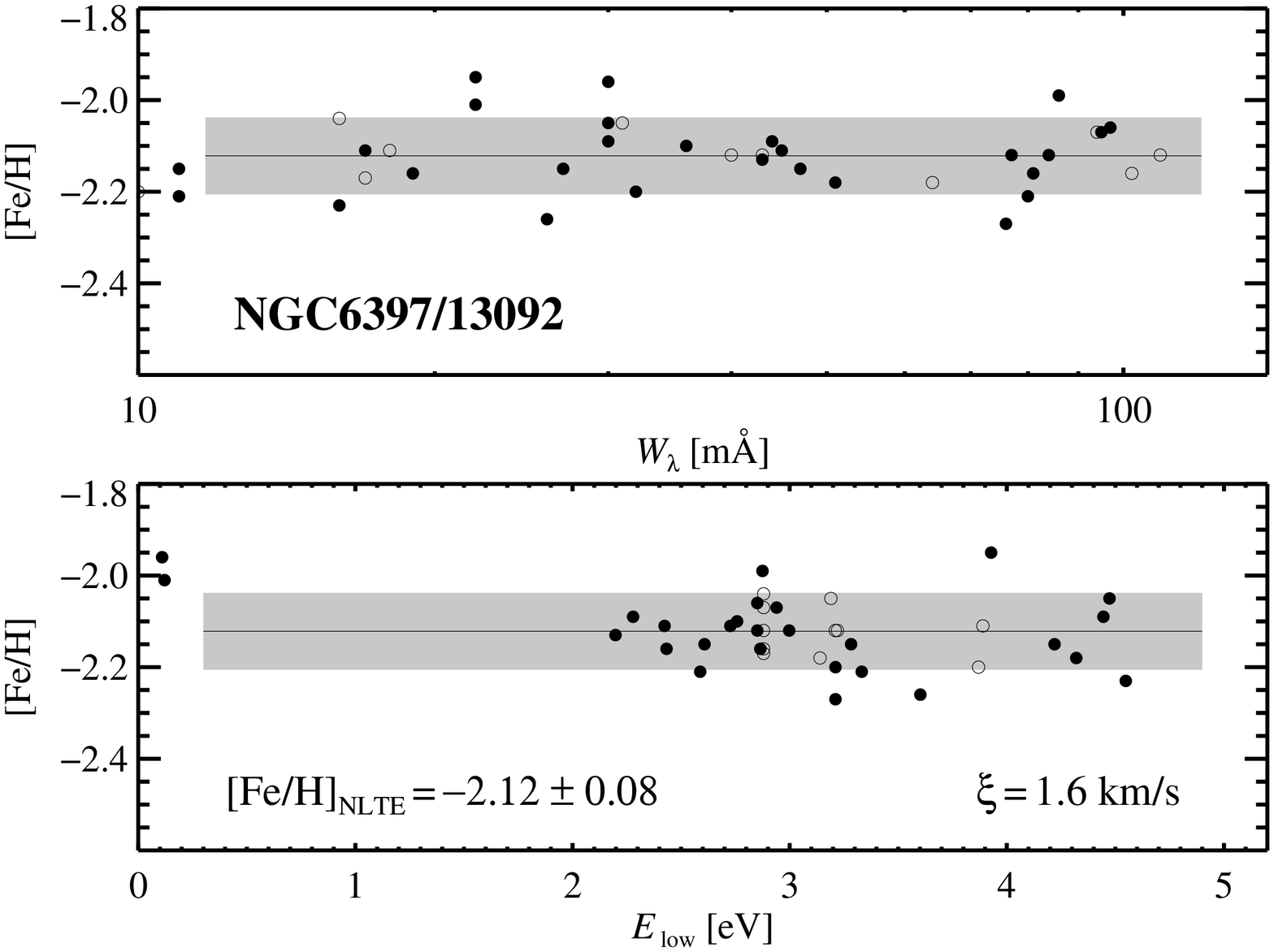}
\caption{Individual line abundances of Fe\,{\sc i} (bullets) and Fe\,{\sc ii} (circles) lines as a function of line strength ({\em top}) and excitation energy of the lower level $E_{\rm low}$ ({\em bottom}) for a TOP star and an RGB star, respectively. While there is no obvious trend of [Fe/H] with $E_{\rm low}$, this is not true for all targets and we do not use the excitation equilibrium as an additional effective-temperature indicator.}\label{micro}
\end{figure}

For consistency with \citet{Korn_etal_2003}, we use the resonance-broadening theory of \citet{Ali_Griem_1966}. While the self-broadening theory of \citet{Barklem_etal_2000} describes the neutral-particle interaction more realistically, the impact on the effective temperatures is small (20-40\,K, as shown by test calculations with MARCS) when the analysis is done differentially to the Sun. This is because the improved self-broadening theory produces stronger theoretical profiles both for the Sun and metal-poor stars analysed here. $\Delta T_{\rm eff}$ is affected by 20\,K ($<$\,0.2\,\%) which is not significant for our conclusions. This also means that broadening-theory differences cannot explain the effective-temperature differences between G01 and this study. A direct comparison of theoretical profiles (R.~Gratton, priv. comm. 2005) indicated that differences in the implementation (model atmosphere, broadening etc.) are of the order of 50\,K.

The effective-temperature determination is done in two steps: firstly, the spectral region of H$\alpha$ is normalized using continuum points at least $\pm$50\,\AA\ away from the line centre. These points are linearly interpolated across the H$\alpha$ profile to obtain a normalized spectrum. Secondly, the normalized H$\alpha$ profile is compared with theoretical profiles by eye to achieve a good overall agreement between theory and observation. The core of H$\alpha$ (the innermost $\pm$ 2\,\AA, residual flux below 0.75) is disregarded.

Figure \ref{Halpha} shows the spectrum of TOP 507433 (G01's target 201432) around H$\alpha$. The UVES spectrum (top, kindly made available by E.~Carretta) shows echelle-blaze and order-merging residuals with an amplitude of $\pm$2\,\% which limit the effective-temperature determination to an accuracy of $\pm$200\,K. However, the error associated with the residual order pattern is likely to have a systematic effect on the effective-temperature determination, as the Balmer-line wings are systematically suppressed (H$\alpha$ falls into the convex part of this pattern). Thus, as overestimation of $T_{\rm eff}$ by as much as 200\,K appears possible \citep{Korn_etal_2004}. This conclusion is supported by the FLAMES-UVES spectrum (middle) of the same star which does not seem to suffer from these instrumental artefacts. An effective temperature of 6260\,K is derived, a value that also finds support from the analysis of the GIRAFFE-MEDUSA spectrum (bottom, $R$\,=\,29\,000, FSR $\geq$ 200\,\AA) of this star (Lind et al., in preparation). A careful re-analysis of the archival data of G01 for this star using the REDUCE software package \citep{Piskunov_Valenti_2002} pointed towards an effective temperature of $T_{\rm eff}$\,=\,6230\,K, again in line with the new spectra obtained.

We note that the problems encountered with the UVES spectra have a far smaller influence on the effective-temperature determination of the bRGB (and RGB) stars, as their Balmer-line profiles are significantly narrower.

If we assume an uncertainty for deriving $T_{\rm eff}$ for an individual star to be 200\,K, then $\Delta T_{\rm eff}$\,(TOP $-$ RGB) turns out to be 1124\,K $\pm$ 120\,K (see Table~\ref{stellparam}). This value is in excellent agreement with the majority of the photometric indices presented in Table~\ref{table:phot} and discussed above.

\subsubsection{Surface gravities}
The iron ionization equilibrium is used to constrain log\,$g$ via the requirement that Fe\,{\sc i} and Fe\,{\sc ii} lines show the same mean abundance. Fe\,{\sc i} is modelled in non-LTE, using the photo-ionization cross-sections computed by \citet{Bautista_1997}, modelling in detail the UV fluxes and treating inelastic collisions with hydrogen according to \citet{Drawin_1968} and \cite{Steenbock_Holweger_1984} with a scaling factor of 3 (for details see \cite{Korn_etal_2003} and references therein). At the same time, the microturbulence is derived by requiring that Fe\,{\sc i} and Fe\,{\sc ii} line abundances show no trend with line strength.

The iron ionization equilibrium is considered as established if the difference in the mean abundances derived from Fe\,{\sc i} and Fe\,{\sc ii} is 0.01\,dex or less. This means that we determine log\,$g$ values with an accuracy of 0.03\,dex.

It is worth noting that overionization in Fe\,{\sc i} is found to be only 0.03\,--\,0.05\,dex for all stars (this is because the calibrated non-LTE model employs rather efficient hydrogen collisions). This means that the absolute log\,$g$ values are raised by about 0.1\,dex for all stars compared to the LTE case. There are practically no differential non-LTE effects. In this sense, our conclusions are not affected by particular choices as regards the underlying line-formation theory.

$\Delta \log g$ is determined to be 1.33 $\pm$ 0.12 (assuming an uncertainty of 0.2\,dex in log $g$ for the individual star, see Table~\ref{stellparam}). This compares very favourably with the photometric estimates based on $\Delta V$ of $\Delta \log g_{\rm phot}$\,=\,1.38 $\pm$ 0.05.

We note that some authors find smaller microturbulence values for TOP stars. For example, G01 assign a common microturbulence of $\xi$\,=\,1.32\,km/s to their TOP stars. With a mean $\xi$ of 2\,km/s, our TOP-star analyses give systematically higher values. However, these values are only marginally (0.2\,km/s) higher than value derived for local TOP star HD 84937 ($T_{\rm eff}$\,=\,6350\,K, log $g$\,=\,4.0, [Fe/H]\,=\,$-$2.16, $\xi$\,=\,1.8\,km/s) using the same methodology (see \citet{Korn_etal_2003} for details). The higher microturbulence values could be caused by the line selection which in turn is constrained by the wavelength coverage and the S/N. To minimize the impact of this parameter on the overall analysis, we made an effort to base the analysis on lines as weak as possible (see Appendix A). Simultaneously, such weak lines have the highest possible abundance sensitivity. Figure~\ref{micro} give examples of the line abundances as a function of line strength and excitation energy for a TOP and an RGB star.

\begin{table*}[!t]
\caption{Mean stellar parameters and elemental abundances of the 18 stars observed with FLAMES-UVES. }
\label{stellparam2}
\begin{center}
\begin{tabular}{rccccccccc}
\tableline
group & $T_{\rm eff}$ [K] & log $g$ & [Fe/H] & log $\varepsilon$\,(Li)\tablenotemark{a}  & log $\varepsilon$\,(Mg)\tablenotemark{b} & log $\varepsilon$\,(Ca)\tablenotemark{c} & log $\varepsilon$\,(Ti)\tablenotemark{d} & log $\varepsilon$\,(Ba)\tablenotemark{e}\\
 & & & NLTE & LTE & NLTE & NLTE & LTE & NLTE \\
\tableline\tableline
TOP$_{\rm ave}$ & 6254 & 3.89\tablenotemark{f}  & $-$2.28  & 2.24 & 5.64 & 4.52 & 2.99 & $-$0.12\\
SGB$_{\rm ave}$ & 5805 & 3.58\tablenotemark{f}  & $-$2.24  & 2.36 & 5.71 & 4.52 & 2.91 & $-$0.10\\
bRGB$_{\rm ave}$ & 5456 & 3.37  & $-$2.18  & 1.38 & 5.72 & 4.54 & 3.01 & $-$0.08\\
RGB$_{\rm ave}$ & 5130 & 2.56  & $-$2.12  & 0.98 & 5.85 & 4.59 & 3.05 & $-$0.12\\
\tableline
\tablenotetext{a}{Based on Li\,{\sc i} 6707 (see Appendix A)}
\tablenotetext{b}{Based on Mg\,{\sc i} 5528 (see Appendix A)}
\tablenotetext{c}{Based on Ca\,{\sc i} 6122, 6162 and 6439 (see Appendix A)}
\tablenotetext{d}{Based on Ti\,{\sc ii} 5188 and 5226 (see Appendix A)}
\tablenotetext{e}{Based on Ba\,{\sc ii} 6496 (see Appendix A)}
\tablenotetext{f}{Not corrected for helium diffusion (see Section \ref{helium})}
\end{tabular}
\end{center}
\end{table*}
\subsection{Comparison between photometry and spectroscopy}
The best agreement between photometry and spectroscopy is obtained for the Str\"{o}mgren indices $(b-y)$ and $(v-y)$. $(V-I)$ performs as well as $(v-y)$ as regards $\Delta T_{\rm eff}$, but the absolute effective temperatures are offset towards cooler temperatures by roughly 100\,K. $B-V$ indicates a smaller $\Delta T_{\rm eff}$ of 906\,K. The agreement with the spectroscopic $\Delta T_{\rm eff}$ is thus excellent for three out of four colour indices.

Due to the low sensitivity of the photometric $\Delta \log g$ values to $\Delta T_{\rm eff}$, all indices indicate a $\Delta \log g$ of around 1.38 (only $(B-V)$ points towards a value of 1.31). All these values are fully compatible with the spectroscopic value of 1.33 $\pm$ 0.12.

\subsection{Comparison with other studies on NGC 6397}
It now becomes clear why we find abundance trends where G01 reported iron abundances of TOP and bRGB stars to ``agree perfectly'': the difference lies in the effective temperatures assigned to the TOP stars which results in a lower effective-temperature difference relative to other groups of stars (bRGB or RGB stars, see above). For bRGB star 6391 (G01's target 793), we derived $T_{\rm eff}$\,=\,5480\,K confirming the effective-temperature estimate of G01 \citep{Korn_etal_2004}.

The new spectroscopic effective temperatures are judged to be more trustworthy (see Sect. \ref{reduction}). Furthermore, the new effective temperatures find independent support from photometry and from the fact that the iron ionization equilibrium could be established for all stars. Neither of these requirements were included in the G01 analysis\footnote{Local subdwarfs were observed in the same observing program and used as a point of reference for $T_{\rm eff}$. However, none of these has an effective temperature above 6100\,K. As the problems with blaze residuals and order merging worsen towards hotter effective temperatures (broader Balmer line profiles), this poor coverage in $T_{\rm eff}$ may have caused the systematic effects on the TOP-star effective temperatures to go undetected.}

The effective-temperature difference between TOP and bRGB stars is spectroscopically determined to be 798\,K and three of the four photometric indices point towards values between 806\,K and 833\,K. $(B-V)$ predicts a markedly lower effective-temperature difference of 615\,K. All these values are significantly lower than the 1000\,K derived by G01.

As a result of the lower effective-temperature difference between TOP and bRGB stars, the gravity difference also turns out to be lower than that derived by G01. At 0.1\,dex, the difference is, however, not significant.

In addition to the re-ana\-lyses of G01 stars, we also analyse one of the TOP stars observed by \citet{Thevenin_etal_2001}. This star (A2084) has the same $V$ magnitude as NGC 6397/12318. Based on the UVES pipeline output (kindly made available by F.~Th\'{e}venin), we confirm the stellar parameters and iron abundance derived by \citet{Thevenin_etal_2001} to within $1\sigma$ measurement uncertainties: $T_{\rm eff}$ = 6250\,K $\pm$ 200\,K, log $g$ = 4.0 $\pm$ 0.2, [Fe/H] = $-$2.27 $\pm$ 0.1, $\xi$ = 1.7\,km/s $\pm$ 0.2\,km/s. This indicates once more that our TOP-star effective temperatures are essentially correct.

\subsection{Chemical abundances}\label{chemabund}
It may seem that one has a great variety of elements to choose from when investigating atomic-diffusion predictions like those shown in Figure~\ref{predictions}. However, various groups of elements are not suited for testing such predictions by comparing their abundances in unevolved and evolved globular-cluster stars. Helium is unobservable in these cool stars. Lithium can be observed in TOP stars, but is significantly processed in RGB stars. Beryllium and boron require near-UV spectroscopy and suffer from the same type of processing as lithium. Carbon and nitrogen in RGB stars may rather reflect the contamination of the atmosphere with material processed in the stellar core. Oxygen, sodium, magnesium and aluminium suffer from the anti-correlations discussed in the Introduction. We are then left with elements like silicon, sulphur, calcium, titanium and iron-group elements (no predictions currently exist for elements heavier than nickel).

According to Figure~\ref{predictions}, a clear signature of atomic diffusion and mixing would be steep trends ($\Delta \log \varepsilon$\,(X) = $ \log \varepsilon$\,(X)$_{\rm RGB} - \log \varepsilon$\,(X)$_{\rm TOP}$ = 0.2) in silicon, rather flat trends ($\Delta \log \varepsilon$\,(X) = 0.1) in calcium and titanium and intermediate trends ($\Delta \log \varepsilon$\,(X) = 0.15) in iron-group elements.

Once the stellar parameters, and in particular stellar-parameter differences, are determined it is a straightforward task to derive abundances, and potential abundance differences, for various elements. As abundance differences are a differential indicator, the exact choice of atomic data for a given line is not critical. This is why we do not tabulate the line data in detail (the lines used are specified in Appendix A). However, even differential indicators may be affected by departures from local thermodynamic equilibrium (LTE), as excitation, ionization and collisions are inherently different in dwarf and giant stars. We therefore made an effort to model as many elements as possible using non-LTE line formation: Mg\,{\sc i} is modelled following \citet{Gehren_etal_2004}, Ca\,{\sc i} following \citet{Mashonkina_etal_2007}, Fe\,{\sc i} following \citet{Korn_etal_2003} and Ba\,{\sc ii} following \citet{Mashonkina_etal_1999}. Lines arising from transition in Ti\,{\sc ii} and Fe\,{\sc ii} are believed to be formed under near-LTE conditions; these species constitute the dominant ionization stages of the respective element. The case of Li\,{\sc i} is discussed in Section~\ref{lithium}.

After co-adding the spectra within each group of stars, lines as weak as 20\,m\AA\ could easily be measured in the TOP stars. Such lines (depending on the line-specific atomic data) have line strengths of 60\,-\,70\,m\AA\ in the RGB stars. The analysis is then insensitive to $\xi_{\rm TOP}$, but has some sensitivity to $\xi_{\rm RGB}$. In Table~\ref{table:errors}, the sensitivity of a few representative lines to variations in stellar parameters (including microturbulence) is given.

\begin{figure*}[!t]
\includegraphics[angle=0,scale=1.35]{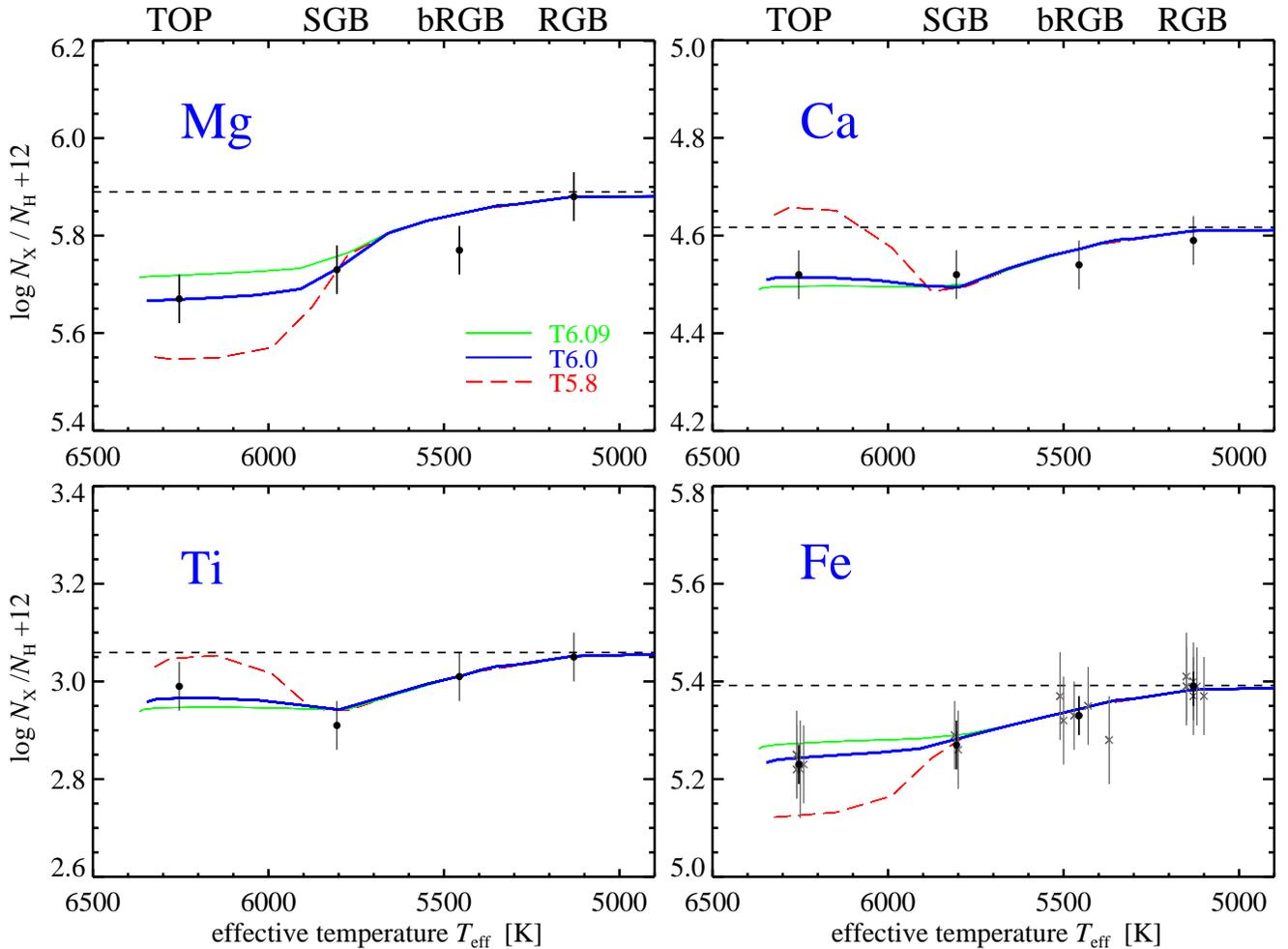}
\caption{Observed trends of elemental abundances with evolutionary stage (bullets with error bars) compared to predictions from stellar-evolution models including atomic diffusion and turbulent mixing with three different efficiencies. Note how the T5.8 model falls below or above the best-fitting T6.0 model depending on the element (likewise the T6.09 model). For details, see text.}
\label{trends}
\end{figure*}

We find systematic trends of abundance with evolutionary phase for iron and magnesium. Propagating the standard deviation of all iron lines into the mean iron abundance of each group of stars, the abundance difference $\Delta \log \varepsilon$\,(Fe) = 0.16 $\pm$ 0.05 is significant at the 3$\sigma$ level. Likewise, assuming an uncertainty of 0.05\,dex on the magnesium abundance determined from the weak Mg\,{\sc i} 5528 line, the {\em relative} abundance trend $\Delta \log \varepsilon$\,(Mg) = 0.21 $\pm$ 0.07 is also significant at the 3$\sigma$ level. As shown in Table~\ref{stellparam2}, calcium, titanium and barium show shallower trends of weak statistical significance.

\section{Comparison with diffusion models} \label{compdiff}
The atomic-diffusion models were computed as described in \cite{Richard_etal_2005} and references therein for the metallicity of NGC~6397.
The abundances of $\alpha$ elements were increased by 0.3\,dex compared to the solar mixture.
These models take into account atomic diffusion (gravitational settling, radiative acceleration, thermal diffusion,
diffusion due to concentration gradients) in a self-consistent way. The detailed treatment of atomic diffusion is
described in \citet{TuRiMiIgRo98} and the radiative accelerations are from \citet{TuRiMiIgRo98} with corrections for
redistribution from \citet{GoLeArMi95} and \citet{LeMiRi2000}. The Rosseland opacity and radiative accelerations are
computed at each time step in each layer for the exact local chemical composition using OPAL monochromatic opacities
for 24 elements. Convection and semi-convection are modelled as diffusive processes as described in \citet{RiMiTu2000} and \citet{RiMiRi2001}.

The main free parameter in these calculations is the efficiency of (turbulent) mixing below the outer convection zone. As discussed in \citet{Richard_etal_2005}, this extra ingredient is needed to meet the constraints imposed by the observations of $^7$Li in warm halo stars: a thin and flat plateau can only be retained in the presence of diffusion, if extra mixing is introduced into the models.

As no parameter-free physical description of turbulent mixing is available, it is introduced into the stellar-evolution models in an ad-hoc manner. The turbulent diffusion coefficient, $D_T$, is linked to the atomic-diffusion coefficient of He at a reference temperature $T_0$ and is chosen to vary with density $\rho$ to the inverse third power:
\begin{displaymath}
D_T = 400 \,D_{\rm He}(T_0)(\frac{\rho}{\rho(T_0)})^{-3}
\end{displaymath}
There are two tunable parameters in this formula, the proportionality constant and the exponent. A steep $\rho$ dependence, i.e. localizing this mixing to a narrow region below the outer convection zone, is suggested by the solar beryllium abundance which is believed to be essentially unaltered since the formation of the Solar System \citep{Proffitt_Michaud_1991}. We thus keep this parameter fixed to $-3$. The assumed overall strength of turbulent mixing can be changed by modifying the proportionality constant or the reference temperature $T_0$. Following \citet{Richard_etal_2005}, the proportionality constant is kept fixed and we vary the reference temperature between log $T$ = 5.8 and 6.09. This is the range of least efficient turbulent mixing that is compatible with the constraints of observed lithium abundances in warm halo stars.

In Figure~\ref{trends}, the observed abundance trends are compared with predictions from stellar evolution models. We find the T6.0 model to match our observations best. Higher-efficiency mixing produces shallower trends for most elements. However, calcium and titanium can be levitated by radiative forces and additional mixing suppresses this tendency making trends mildly steeper. The element-specific trends observed and their agreement with the atomic-diffusion model suggest a causal connection between the abundance trends and effects due to atomic diffusion moderated by mixing below the convective envelope. We have empirically determined the amount of mixing that is required to reproduce the observed trends.

\subsection{Error analyses}\label{errors}
\begin{table*}
\caption{Influence of stellar parameters on abundances for a few representative lines. These are computed using MARCS by matching group-averaged equivalent widths.\label{table:errors}}
\begin{center}\small
\begin{tabular}{lrrrrrrr}
\tableline\tableline
variation   &      Li I 6707 &   Mg I 5528 &   Ca I 6122 &   Ti II 5226  &  Fe II 5018  &  Fe II 5234  &  Fe I 6494\\
\tableline
TOP\\
  $T_{\rm eff}$ + 100\,K &     0.070  &    0.047  &    0.064  &   0.035  &    0.022  &   0.011   &   0.078\\
  log $g$ + 0.3\,dex &    $-$0.002  &   $-$0.011  &   $-$0.006  &   0.100  &    0.091  &   0.105   &  $-$0.003\\
   $\xi$ + 0.3\,km/s &    $-$0.003  &   $-$0.010  &   $-$0.009  &  $-$0.005  &   $-$0.089  &  $-$0.005   &  $-$0.010\\
  $T_{\rm eff}$ + 200\,K &     0.139  &    0.089  &    0.122  &   0.069  &    0.042  &   0.023   &   0.154\\
  log $g$ + 0.6\,dex &    $-$0.003  &   $-$0.024  &   $-$0.017  &   0.200  &    0.174  &   0.210   &  $-$0.004\\
   $\xi$ + 0.6\,km/s &    $-$0.004  &   $-$0.021  &   $-$0.020  &  $-$0.022  &   $-$0.158  &  $-$0.002   &  $-$0.019\\
  $T_{\rm eff}$ $-$ 100\,K &    $-$0.072  &   $-$0.041  &   $-$0.055  &  $-$0.036  &   $-$0.023  &  $-$0.011   &  $-$0.080\\
  log $g$ $-$ 0.3\,km/s &     0.003  &    0.016  &    0.014  &  $-$0.100  &   $-$0.096  &  $-$0.104   &   0.003\\
   $\xi$ $-$ 0.3\,km/s &    $-$0.003  &    0.017  &    0.020  &   0.010  &    0.113  &   0.010   &   0.019\\
  $T_{\rm eff}$ $-$ 200\,K &    $-$0.151  &   $-$0.087  &   $-$0.117  &  $-$0.073  &   $-$0.047  &  $-$0.023   &  $-$0.162\\
  log $g$ $-$ 0.6\,dex &     0.008  &    0.025  &    0.022  &  $-$0.200  &   $-$0.197  &  $-$0.207   &   0.007\\
   $\xi$ $-$ 0.6\,km/s &    $-$0.003  &    0.030  &    0.036  &   0.027  &    0.249  &   0.020   &   0.024\\
SGB\\
  $T_{\rm eff}$ + 100\,K &     0.079  &    0.055  &    0.075  &   0.040  &    0.030  &   0.013   &   0.091\\
  log $g$ + 0.3\,dex &    $-$0.000  &   $-$0.027  &   $-$0.019  &   0.099  &    0.069  &   0.104   &  $-$0.008\\
   $\xi$ + 0.3\,km/s &    $-$0.010  &   $-$0.017  &   $-$0.019  &  $-$0.017  &   $-$0.128  &  $-$0.010   &  $-$0.028\\
  $T_{\rm eff}$ + 200\,K &     0.155  &    0.107  &    0.142  &   0.078  &    0.059  &   0.025   &   0.178\\
  log $g$ + 0.6\,dex &    $-$0.000  &   $-$0.056  &   $-$0.045  &   0.197  &    0.128  &   0.206   &  $-$0.018\\
   $\xi$ + 0.6\,km/s &    $-$0.020  &   $-$0.033  &   $-$0.040  &  $-$0.032  &   $-$0.231  &  $-$0.013   &  $-$0.051\\
  $T_{\rm eff}$ $-$ 100\,K &    $-$0.081  &   $-$0.052  &   $-$0.067  &  $-$0.041  &   $-$0.033  &  $-$0.013   &  $-$0.094\\
  log $g$ $-$ 0.3\,km/s &     0.001  &    0.029  &    0.025  &  $-$0.099  &   $-$0.079  &  $-$0.103   &   0.006\\
   $\xi$ $-$ 0.3\,km/s &     0.009  &    0.021  &    0.032  &   0.021  &    0.148  &   0.016   &   0.034\\
  $T_{\rm eff}$ $-$ 200\,K &    $-$0.165  &   $-$0.110  &   $-$0.141  &  $-$0.083  &   $-$0.067  &  $-$0.026   &  $-$0.191\\
  log $g$ $-$ 0.6\,dex &     0.003  &    0.050  &    0.040  &  $-$0.198  &   $-$0.166  &  $-$0.206   &   0.011\\
   $\xi$ $-$ 0.6\,km/s &     0.017  &    0.040  &    0.060  &   0.047  &    0.301  &   0.034   &   0.078\\
bRGB\\
  $T_{\rm eff}$ + 100\,K &     0.088  &    0.064  &    0.092  &   0.046  &    0.036  &   0.016   &   0.111\\
  log $g$ + 0.3\,dex &    $-$0.002  &   $-$0.053  &   $-$0.042  &   0.094  &    0.051  &   0.105   &  $-$0.029\\
   $\xi$ + 0.3\,km/s &    $-$0.003  &   $-$0.027  &   $-$0.031  &  $-$0.035  &   $-$0.137  &  $-$0.013   &  $-$0.063\\
  $T_{\rm eff}$ + 200\,K &     0.172  &    0.129  &    0.176  &   0.105  &    0.075  &   0.030   &   0.218\\
  log $g$ + 0.6\,dex &    $-$0.003  &   $-$0.099  &   $-$0.092  &   0.187  &    0.092  &   0.198   &  $-$0.065\\
   $\xi$ + 0.6\,km/s &    $-$0.005  &   $-$0.051  &   $-$0.064  &  $-$0.063  &   $-$0.251  &  $-$0.029   &  $-$0.114\\
  $T_{\rm eff}$ $-$ 100\,K &    $-$0.091  &   $-$0.073  &   $-$0.083  &  $-$0.048  &   $-$0.040  &  $-$0.012   &  $-$0.116\\
  log $g$ $-$ 0.3\,km/s &     0.003  &    0.046  &    0.050  &  $-$0.096  &   $-$0.067  &  $-$0.099   &   0.023\\
   $\xi$ $-$ 0.3\,km/s &     0.002  &    0.021  &    0.048  &   0.043  &    0.150  &   0.027   &   0.079\\
  $T_{\rm eff}$ $-$ 200\,K &    $-$0.187  &   $-$0.144  &   $-$0.176  &  $-$0.096  &   $-$0.081  &  $-$0.023   &  $-$0.236\\
  log $g$ $-$ 0.6\,dex &     0.006  &    0.087  &    0.089  &  $-$0.193  &   $-$0.134  &  $-$0.200   &   0.039\\
   $\xi$ $-$ 0.6\,km/s &     0.004  &    0.044  &    0.091  &   0.097  &    0.292  &   0.056   &   0.174\\
RGB\\
  $T_{\rm eff}$ + 100\,K &     0.099  &    0.070  &    0.104  &   0.051  &    0.040  &   0.018   &   0.127\\
  log $g$ + 0.3\,dex &    $-$0.007  &   $-$0.074  &   $-$0.066  &   0.087  &    0.041  &   0.098   &  $-$0.044\\
   $\xi$ + 0.3\,km/s &    $-$0.002  &   $-$0.054  &   $-$0.073  &  $-$0.095  &   $-$0.178  &  $-$0.039   &  $-$0.123\\
  $T_{\rm eff}$ + 200\,K &     0.193  &    0.144  &    0.202  &   0.101  &    0.079  &   0.033   &   0.248\\
  log $g$ + 0.6\,dex &    $-$0.014  &   $-$0.143  &   $-$0.142  &   0.170  &    0.074  &   0.185   &  $-$0.099\\
   $\xi$ + 0.6\,km/s &    $-$0.003  &   $-$0.098  &   $-$0.140  &  $-$0.166  &   $-$0.342  &  $-$0.073   &  $-$0.221\\
  $T_{\rm eff}$ $-$ 100\,K &    $-$0.103  &   $-$0.087  &   $-$0.102  &  $-$0.051  &   $-$0.041  &  $-$0.008   &  $-$0.134\\
  log $g$ $-$ 0.3\,km/s &     0.008  &    0.054  &    0.065  &  $-$0.091  &   $-$0.051  &  $-$0.092   &   0.036\\
   $\xi$ $-$ 0.3\,km/s &     0.002  &    0.039  &    0.085  &   0.126  &    0.172  &   0.060   &   0.149\\
  $T_{\rm eff}$ $-$ 200\,K &    $-$0.212  &   $-$0.167  &   $-$0.212  &  $-$0.099  &   $-$0.079  &  $-$0.016   &  $-$0.275\\
  log $g$ $-$ 0.6\,dex &     0.018  &    0.105  &    0.117  &  $-$0.179  &   $-$0.111  &  $-$0.190   &   0.064\\
   $\xi$ $-$ 0.6\,km/s &     0.004  &    0.082  &    0.164  &   0.277  &    0.313  &   0.129   &   0.311\\
\tableline
\end{tabular}
\end{center}
\end{table*}

Here we address the question of whether the observed abundance trends are caused by contrived sets of stellar parameters or fundamental assumptions made in the modelling of stellar atmospheres, i.e.~the hydrostatic approximation with convection according to mixing-length theory.

\subsubsection{Stellar parameters}
Given that G01 found no abundance differences in iron between their TOP and bRGB stars, we can ask ourselves whether all abundance trends vanish if we assume TOP effective temperatures $\approx$ 250\,K hotter. As can be seen from Table~\ref{table:errors}, this is not the case. An extreme example is the trend derived from Mg\,{\sc i} 5528 which would require well above 400\,K hotter TOP stars (or 300\,K cooler RGB stars) to completely remove the abundance trend. Even larger $T_{\rm eff}$ corrections would be required, if one were to consider the feedback from log\,$g$. Given that $\Delta T_{\rm eff}$ is around 1100\,K, such corrections are of the order of 30\,\%. We regard it as unlikely that we underestimate $\Delta T_{\rm eff}$ by as much as one third.

Along the same lines, the diffusion signature in lithium between the TOP and SGB stars of 0.12\,dex would require 170\,K hotter TOP stars or 150\,K cooler SGB stars. The effective-temperature difference between these two groups of stars is no larger than 450\,K, thus a correction of more than 30\,\% would also be required here.

When it comes to log $g$-sensitive species like Fe\,{\sc ii}, surface-gravity corrections of $\geq$0.4\,dex are required to harmonize TOP and RGB iron abundances ($\Delta \log g$ would then be larger than 1.7). Given that $\Delta \log g_{\rm phot}$ is robustly determined from $\Delta V$, a $\Delta \log g$ larger than 1.5 can practically be ruled out.

In the absence of spectroscopic or photome\-tric evidence for either hotter TOP-star or cooler RGB-star effective temperatures, it seems rather unlikely that these trends are produced by biased stellar parameters. Increasing TOP-star effective temperatures by 100\,K to remove the abundance trends, as claimed by \citet{Bonifacio_etal_2007}, is not possible. It is one of the strengths of this analysis that it combines independent  techniques to derive stellar parameters which turn out to agree rather well.

\subsubsection{Model atmospheres}
\subsubsection*{Effects of the first dredge-up}
The giant stars of NGC 6397 are expected to be affected
by the first dredge-up of CN processed material, and more so the higher
up the stars are along the giant branch. Along the lower giant branch
the Pop II stars are predicted to be only marginally affected, corresponding to
a decrease of [C/Fe] by about 0.15\,dex or less \citep{Charbonnel_1994, Vandenberg_Smith_1988} which has also been verified by \citet{Gratton_etal_2000} for Pop II field stars, and a corresponding increase of [N/Fe] keeping the
absolute number of C+N nuclei constant. A much more severe decrease of [C/Fe]
matched by an [N/Fe] increase is found by Gratton et al. (2000) for field
stars that have evolved off the bump in the RGB luminosity function around log
$L\,=\,1.8\,L_\odot$ where a reduction by a factor of 2.5 in the carbon abundance is
common. For cluster giants, even deeper mixing may occur which may contribute to the
O-depletion/Na-enhancement observed \citep{Kraft_1994}. A question is now
whether such phenomena which vary systematically from the TOP up along the RGB
may cause differential effects on the stellar atmospheres that could affect the
measures of abundance differences between our different stellar groups.

Experiments with varying the CNO abundances in a set of MARCS models show that the differential effects in the model structures by such changes
are only marginal, leading to maximum changes in the temperature gradient of $\delta T/\delta \tau$(Ross) of less than 10\,K. Moreover, the gradient is diminished, causing weaker, sooner than stronger, metal lines for the giants than in the
case of unmodified abundances. Thus, we conclude that the observed trends of Fig.~\ref{trends} can not be explained by such dredge-up effects.

\subsubsection*{3D corrections}
\begin{table*}
\caption{3D$-$1D LTE corrections to  Mg, Ti, and Fe abundances
for the metal-poor RGB  and TOP star. Stellar parameters for the models are: $T_{\rm eff}$\,=\,5035\,K, $\log g$\,=\,2.20, [Fe/H]\,=\,$-$2.0, $\xi$\,=\,1.6,km/s (RGB), $T_{\rm eff}$\,=\,6180\,K, $\log g$\,=\,4.04, [Fe/H]\,=\,$-$2.0, $\xi$\,=\,2.0 (TOP).\label{table:3Dcorr} }
\begin{center}
\begin{tabular}{lcrcrr}
\tableline\tableline
ion & $\lambda$ & $\log gf$ & $E_{\rm low}$ & $\log \varepsilon_{\rm{3D}} - \log \varepsilon_{\rm{1D}}$ & $\log \varepsilon_{\rm{3D}} - \log \varepsilon_{\rm{1D}}$ \\
 & [\AA] & & [eV] & RGB & TOP \\
\tableline
Mg\,{\sc i} & 5528 & $-0.620$ & 4.346 & $-0.09$ & $+0.03$ \\
Ti\,{\sc ii} & 5226 & $-1.230$ & 1.566 & $+0.06$ & $+0.01$ \\
Fe\,{\sc i} & 4920 & $ 0.068$ & 2.832 & $-0.08$ & $-0.01$ \\
Fe\,{\sc i} & 5328 & $-1.466$ & 0.915 & $-0.29$ & $-0.45$ \\
Fe\,{\sc i} & 5424 & $ 0.580$ & 4.320 & $-0.02$ & $+0.04$ \\
Fe\,{\sc ii} & 4923 & $-1.260$ & 2.891 & $+0.23$ & $+0.34$ \\
Fe\,{\sc ii} & 5234 & $-2.230$ & 3.221 & $+0.16$ & $+0.08$ \\
Fe\,{\sc ii} & 5316 & $-1.850$ & 3.153 & $+0.26$ & $+0.13$ \\
\tableline
\end{tabular}
\end{center}
\end{table*}

Recent stellar surface convection simulations indicate that the structural differences between 3D hydrodynamic and 1D hydrostatic model atmospheres may have a significant effect on the predicted strength of spectral lines and hence on
the derived spectroscopic abundances in metal-poor stars \citep{Asplund_etal_1999,Asplund_Garcia_Perez_2001,Collet_etal_2006}.
We use here the surface convection simulations of a red giant \citep{Collet_etal_2007} and of a turnoff star \citep{Asplund_Garcia_Perez_2001} at a metallicity of $\rm{[Fe/H]}=-2$ as time-dependent 3D hydrodynamic model atmospheres to calculate Mg, Ti, and Fe spectral lines
and estimate the impact of 3D models on the determination of abundance trends for these elements with respect to stellar evolutionary stage. Corresponding calculations were not made for Ca, since LTE was found to be inappropriate (see Appendix A). The relevant stellar parameters of the 3D model atmospheres considered here are given in the caption of Table~\ref{table:3Dcorr}.

From the full RGB- and TOP-star simulations, we select two representative sequences, respectively, 10000 and 60 minutes
long, of about 40 snapshots taken at regular intervals in time. We then compute spectral line profiles for a number of Mg, Ti, and Fe
lines (see Table~\ref{table:3Dcorr}) under the assumption of LTE. For each line, we solve the radiative transfer equation along 17 directions (four $\mu$-angles, four $\phi$-angles, plus the vertical), after which we perform a disk integration and a time average over all snapshots.

Using the 3D models, we derive the abundances of Mg, Ti, and Fe from the measured equivalent widths of individual lines of these elements.
We then carry out a differential comparison with the results of the abundance analysis based on classical 1D LTE
plane-parallel hydrostatic {\sc MARCS} model atmospheres constructed with the same stellar parameters and chemical compositions as the hydrodynamic simulations. We caution that this 3D-1D analysis is restricted in two ways. Firstly, the analysis is based on equivalent widths and an analysis based on line profiles may return somewhat different results. Secondly, microturbulence enters on the 1D side of the analysis making the results dependent on microturbulence when strong lines (Fe\,{\sc i} 4920, Fe\,{\sc i} 5328 and Fe\,{\sc ii} 4923) are investigated.

The results of the 3D$-$1D differential Mg, Ti, and Fe abundance analysis for the RGB and TOP cases are shown in Table~\ref{table:3Dcorr}. No clear picture emerges: $\Delta \log \varepsilon$\,(Mg) becomes smaller, $\Delta \log \varepsilon$\,(Ti) steeper, while weak and strong Fe\,{\sc i} and Fe\,{\sc ii} lines show corrections modifying $\Delta \log \varepsilon$\,(Fe) between +0.11 and $-$0.16. The average $\Delta \log \varepsilon$\,(Fe)$_{\rm 3D}$ is 0.02\,dex steeper than its 1D counterpart. Judging from the most reliable weak Fe\,{\sc ii} lines, a steeper abundance trend in iron  seems possible (by about 0.1\,dex). However, this does not fit to the 3D trends in Mg and Ti which become flatter and steeper, respectively (see the T6.09 model in Figure~\ref{trends}).
The most important conclusion is that the abundance trends as such are probably not an artefact of the modelling in 1D.

  That 1D flux-constant models are not fully adequate for the
  detailed analysis of metal-poor stars has been realized for some
  time. One could ask whether inaccuracies in the models could mimic
  the observed abundance trends. By experimenting with ad-hoc steepenings of
  the temperature gradients in MARCS models, we have found that a
  lowering of the temperature by typically 400\,K in the outer atmosphere for the giants, keeping the temperature in the
  continuum-forming layers fixed and with applying no corresponding
  steepening for the TOP stars, would remove abundance trends similar
  to those shown in Fig.~6 (see also \cite{Gustafsson_1983}). (We note
  is passing that the Fe\,{\sc i/ii} balance would be affected by such a
  steepening since the effects are smaller for Fe\,{\sc ii} lines. The
  consistency of our ionization balance with the log $g$ values
  derived from the luminosities thus suggests that the trends in
  Fig.~6 are not due to such a steepening. This is, however, an
  uncertain argument, as it is dependent on the correctness of the
  non-LTE calculations for iron.) The steepening of the temperature
  gradient needed is considerable, more than twice the total
  line-blanketing effect on the temperature gradient of the MARCS
  models. It may, however, not be totally excluded. The upper
  temperature in real stars is set by the balance between the
  adiabatic cooling of the up-streaming gas and the heating by
  radiation from below, and the purely adiabatic temperature gradient is below the steepened
  MARCS model. The radiative heating may be overestimated in 1D
  models, and also in present-day 3D ones, if the thermal coupling to
  the gas is lower than our LTE models predict. This could be case if
  scattering processes dominate.

Future analysis in 3D, with a more detailed treatment of the radiative
field, based on line profiles and preferably with stellar parameters derived in a self-consistent way, may change details. However, the
general results of the present study will likely remain.

\subsubsection{Non-LTE effects}
While we made an effort to treat the line formation of the investigated elements as detailed as possible, there are uncertainties in present-day statistical-equilibrium calculations. These mainly arise from poorly-known cross-sections for inelastic collisions with neutral hydrogen. All non-LTE studies listed in Sect.~\ref{chemabund} take an empirical approach in calibrating the efficiency of such collisions on stars with known stellar parameters, i.e.\ local subdwarfs with accurate HIPPARCOS parallaxes. In all cases, it is found that the neglect of hydrogen collisions leads to non-LTE effects that are too large (judged on line-to-line scatter or equilibria of two ionization stages, where available). Larger non-LTE effects are possible to obtain (and could alter the observed abundance trends, if they introduce differential effects), but this would be in conflict with the empirical findings. It should be noted that the non-LTE approach flattens abundance trends for two elements, calcium and barium (see Appendix A). We thus regards it as unlikely that modelling deficits in the current non-LTE treatment conspire to produce the element-specific trends seen in Fig.~\ref{trends}.

\section{Discussion} \label{disc}
In the following, consequences of the atomic-diffusion interpretation of the observed abundance trends are discussed.
\subsection{Helium} \label{helium}

\begin{figure}[!t]
\includegraphics[angle=0,scale=0.50]{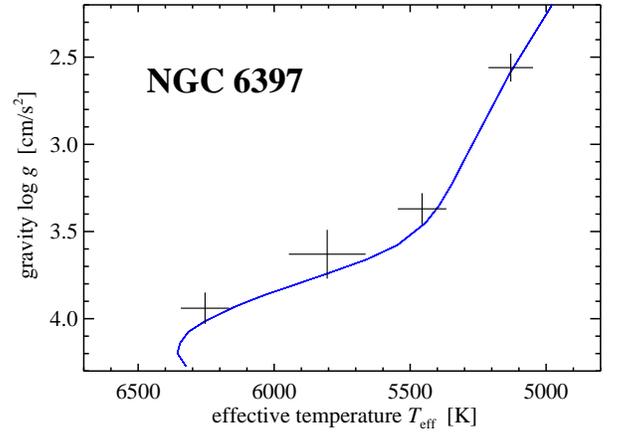}
\caption{Stellar parameters compared to a 13.5\,Gyr isochrone constructed from the T6.0 diffusion model. Helium-diffusion corrections to log $g$ have been applied to the TOP and SGB stars (+0.05\,dex in both cases). The TOP stars have effective temperatures about 100\,K lower than the turn-off point, in line with expectations from photometry. The cluster may be somewhat younger than 13.5\,Gyr which, however, has only a small effect on our comparison between observed abundance trends and atomic-diffusion predictions.}
\label{iso}
\end{figure}

The uncovered abundance trends also have structural effects on the atmosphere which one needs to consider. Due to its high abundance, the downward diffusion of helium (by about 0.25\,dex in the T6.0 model) is accompanied by a corresponding change in mean molecular weight. According to \citet{Stromgren_etal_1982}, this effect can be mapped as a shift in surface gravity. That is, the line spectrum of a helium-normal atmosphere corresponds to the line spectrum of a helium-poor atmosphere at somewhat higher log $g$. This increases the log $g$ values for the TOP and SGB stars by +0.05\,dex, while the original helium abundance is practically recovered for the bRGB stars and, even more so, RGB stars. This improves the agreement between the spectroscopic and photometric $\Delta \log g$ values even further.

The empirical turbulent-mixing calibration constraints the amount of atmospheric helium diffusion. This has, however, little effect on globular-cluster ages, as the central helium diffusion is unaffected by the amount of turbulent mixing acting just below the convective envelope. We emphasize, however, that the new TOP stellar parameters do not contradict the cosmological age constraint. In other words, in spite of effective temperatures 230\,K cooler than what was derived by G01, NGC~6397 is younger than 13.5\,Gyr (see Fig.~\ref{iso}).

\subsection{Lithium}\label{lithium}
\begin{figure}[!t]
\includegraphics[angle=0,scale=0.44]{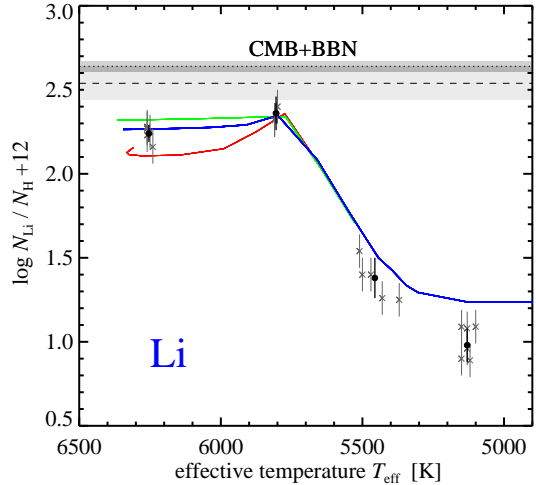}
\caption{Observed trends of lithium. Crosses indicate individual stellar abundances, while bullets shown group averages. Full-drawn lines represent the atomic-diffusion model predictions (cf.~Fig.~\ref{trends}) which have an original abundance as given by the dashed line ($\log \varepsilon$\,(Li)\,=\,2.54 $\pm$ 0.1). The dark-shaded area shows the predicted primordial lithium abundance (``CMB+BBN'': $\log \varepsilon$\,(Li)\,=\,2.64 $\pm$ 0.03, \citet{Spergel_etal_2007}).}
\label{lithiumtrend}
\end{figure}

Lithium shows abundance trends quite different from any of the other elements discussed above. This is because one sees its surface abundance become diluted as the stars evolve to become red giants. The dilution seems to set in just below the SGB-star group, that is, below 5800\,K (see Fig.~\ref{lithiumtrend}). The dilution that the models predict falls successively short of the observed abundances for the bRGB and RGB stars. For the latter group, the discrepancy exceeds 0.2\,dex and is usually ascribed to a source of extra mixing not considered in these calculations \citep{Charbonnel_1995}.

We find lithium abundances in the TOP stars of $\log \varepsilon$\,(Li)\,=\,2.24 $\pm$ 0.05. \citet{Bonifacio_etal_2002} derive lithium abundances for 12 TOP in NGC 6397 drawn from the studies of G01 and \citet{Thevenin_etal_2001}. They derive a mean abundance of $\log \varepsilon$\,(Li)\,=\,2.34 $\pm$ 0.056, 0.1\,dex above our TOP star lithium abundances. For the three stars in common with this study (see Table~\ref{obs}), they assign effective temperatures 145\,K hotter on average. This $T_{\rm eff}$ difference fully explains the offset in lithium abundances.

Comparing the average lithium abundances in the TOP and SGB groups, we detect an abundance difference of 0.12\,dex. The stellar-structure models with atomic diffusion also predict higher abundances in the SGB stars. We have thus identified a diffusion signature on the Spite plateau of lithium \citep{Spite_Spite_1982} that has so far only been identified in a statistical sense \citep{Charbonnel_Primas_2005} in comparing lithium abundances in field dwarf and subgiant stars. With hindsight, this signature may be possibly be traced in the lithium abundances of three TOP stars in NGC 6397 (with $T_{\rm eff}$ values between 6000\,K and 6180\,K) analysed by \citet{Pasquini_Molaro_1996}.

We note that the T5.8 model has turbulent mixing that is too weak to keep the Spite plateau flat in the turn-off region around $T_{\rm eff}$ = 6300\,K. This is indicated in Fig.~\ref{lithiumtrend} by the hook at the hot end of the T5.8 model. It thus seems that our observations constrain  turbulent mixing to a strength that is just sufficient to keep the Spite plateau flat on the main sequence and around the turn-off region.

The atomic-diffusion model predicts the abundance of lithium these NGC 6397 stars had when they formed. This value, $\log \varepsilon$\,(Li)\,=\,2.54 $\pm$ 0.1, compares well with the WMAP-based predictions of standard Big-Bang nucleosynthesis, $\log \varepsilon$\,(Li)\,=\,2.64 $\pm$ 0.03 \citep{Spergel_etal_2007}. It thus seems that atomic diffusion is predominantly responsible for
  photospheric lithium abundances of stars on the Spite plateau being systematically below the primordial value.

In comparing with the primordial lithium abundance, no correction for galacto-chemical production of lithium (by cosmic-ray spallation) is considered. Empirical trends of lithium abundance with metallicity vary from author to author: some find trends as steep as $\approx$ 0.1\,dex per 1\,dex in [Fe/H] \citep{Ryan_etal_1999,Asplund_etal_2006}, others find no trend at all \citep{Melendez_Ramirez_2004,Shi_etal_2007}. The different behaviour is primarily caused by differences in the adopted effective-temperature scale. Accounting for an extra 0.1\,dex lessens the agreement between diffusion-corrected stellar abundances and the primordial value. A gap of 0.1\,--\,0.2\,dex can, however, be explained, e.g., by mixing through Population III stars (see \cite{Piau_etal_2006}, but see also \cite{Prantzos_2006} for some criticism of this scenario).

\subsection{Abundance ratios}
Atmospheric abundances of unevolved, metal-poor stars seem systematically affected by atomic diffusion. This is particularly true for ratios with respect to hydrogen where diffusion corrections can reach 0.2\,--\,0.3\,dex for widely used elements like Mg or Fe (see Fig.~\ref{6397cc}). Abundance ratios of metals (e.g.~Mg/Fe) may only be affected by 0.05\,dex. Nonetheless, for the highest accuracy the predictions from atomic-diffusion models have to be included.

For the time being, we have no understanding of the metallicity dependence of turbulent mixing. It could be that extremely metal-poor stars (like HE 1327-2326, \cite{Frebel_etal_2005}) are therefore more (or less) affected by atomic diffusion. We are in the process of investigating this issue. It is highly desirable to understand the physical processes that give rise to mixing below the convection zone. Encouraging results based on hydrodynamic models have already been presented by \citet{Charbonnel_Talon_2005}.

\subsection{Stellar isochrone ages}
In deriving stellar ages with the help of isochrones, two fundamentally different situations need to be discussed. On the one hand, there are populations of stars (like globular clusters); on the other hand, there are individual field stars. The impact of atomic diffusion on age determinations differs in these two cases, as detailed below.

Globular-cluster ages are unlikely to be affected by our findings, for two reasons: firstly, their metallicities are usually derived from red-giant stars which show in their atmospheres the composition of the gas they formed from (with some notable exceptions). Secondly, helium diffusion has been considered in deriving globular-cluster isochrone ages for quite some time \citep{Chaboyer_etal_1998}.

Ages for unevolved halo field stars will be modified by atomic diffusion. One needs to consider the interplay of metal diffusion (i.e.~comparing to the correct isochrone) and helium diffusion (i.e.~shifting the surface gravity upward). Main-sequence and TOP stars will likely get lower ages, while subgiants may become somewhat older. We postpone a quantitative discussion of these effects to a forthcoming paper.

\subsection{Abundance gradients in the halo}
Large, magnitude-limited surveys like the Sloan Digital Sky Survey (SDSS) provide us with a wealth of information on the chemical composition of the Galactic halo. If one performs analyses of dwarfs and giants in a certain apparent-magnitude range, one inevitably samples different space volumes, more nearby ones with the dwarfs, more distant ones with the giants. Large samples of stars can then define metallicity distribution functions (MDF) and one may want to look for gradients with Galacto-centric distance or height above the Galactic plane. Assume that one finds that the MDFs of distant giants and more nearby dwarfs peak at the same metallicity. Does this mean that there are no abundance gradients in that direction? In the light of atomic diffusion affecting dwarf-star abundances only, this apparent null result actually means that the more distant population of stars is 0.1\,--\,0.2\,dex more metal-poor.

\subsection{Other metallicities}
We recently computed models similar to those used here for [Fe/H]\,= $-$1.5, the metallicity of NGC 6752. The predictions indicate abundance trends practically as large as in NGC 6397 (if turbulent mixing is assumed to be equally efficient). We will study NGC 6752 with FLAMES-UVES and GIRAFFE-MEDUSA to quantify potential effects of atomic diffusion and mixing at this metallicity.

\citet{Cohen_Melendez_2005} studied M 13 at [Fe/H]\,$\approx$\,$-$1.5, but found no abundance trends. However, their work only included three unevolved (subgiant) stars. This study may thus not sample the region where atomic diffusion acts in sufficient detail. Nonetheless, the three least evolved stars show a mean iron abundance 0.15\,dex below the average of the other 22 (giant) stars.

Better sampling of the turnoff region was achieved by \citet{Ramirez_Cohen_2003} for M~5 at [Fe/H]\,$\approx$\,$-$1.3 (six TOP stars). To within the abundance uncertainties (claimed to be 0.1\,dex), they find no trends for a variety of elements. Concerns about the data quality could be raised, as the TOP stars have S/N ratios of 40\,--\,50 {\em per 4-pixel resolution element}. However, there is currently no satisfactory explanation for the absence of abundance trends in M 5, unless one proposes that turbulent mixing is markedly more efficient than the strength found in NGC 6397.

At even higher metallicities, effects are expected to diminish, as the convection zones of more metal-rich stars are more massive than those of metal-poor ones.

\section{Conclusions} \label{conc}
\begin{figure}[!t]
\includegraphics[angle=0,scale=0.39]{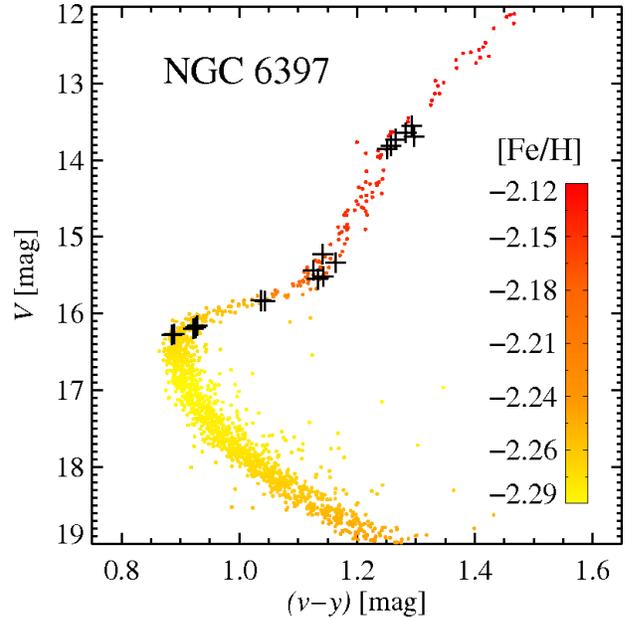}
\caption{Colour-magnitude diagram of NGC 6397 with colour-coding according to the atomic-diffusion model with turbulent mixing as calibrated on the observed stars (crosses).}
\label{6397cc}
\end{figure}

We have identified significant abundance trends for the elements lithium, magnesium and iron between groups of stars from the turnoff point to the bump on the red-giant branch. Other elements studied (calcium, titanium and barium) show shallower trends of weak statistical significance. For lithium, the global trend can be explained by dilution of the atmospheric layers with lithium-free material as the convective envelope expands inward. For the other elements, we show that atomic diffusion (the net effect from gravitational settling and radiative levitation) and turbulent mixing below the convective envelope can describe the trends in an element-specific way. A diffusion signature is also identified for lithium in stars on the Spite plateau and their original lithium abundance is inferred to be $\log \varepsilon$\,(Li)\,=\,2.54 $\pm$ 0.1, in good agreement with WMAP-based predictions of Big-Bang nucleosynthesis \citep{Spergel_etal_2007}.

Our study has consequences for the use of unevolved metal-poor stars as tracers of the chemical evolution of the Galaxy. Their isochrone ages are also systematically affected by atomic diffusion.

For the time being, the turbulent mixing needed to bring theoretical predictions in agreement with observations is prescribed in an ad-hoc way. It would be desirable to isolate the physical process that gives rise to the turbulent mixing below the convective envelope. The predictions of hydrodynamic models will have to be put to the test by means of observations like the ones we presented here.

Recently, $^6$Li was traced in several metal-poor field stars at levels well above those that can be explained by Galactic production \citep{Asplund_etal_2006}. The atomic-diffusion corrections derived here ($\approx$ +0.25\,dex, see Fig.~\ref{predictions}) elevate the abundances even further. Together with pre-main-sequence burning, they thus aggravate the $^6$Li problem. If the detections are confirmed by more sophisticated modelling (but cf.~\cite{Cayrel_2007}), it is unclear where all this $^6$Li was synthesized; various processes are investigated \citep{Jedamzik_etal_2006,Tatischeff_Thibaud_2007}. Lithium thus remains an enigmatic element.

We are only beginning to investigate the effects of atomic diffusion and mixing in old stars observationally. More work is needed to establish their impact on stellar atmospheric abundances and ages. Globular clusters like NGC 6397 (see Fig.~\ref{6397cc}) will play a key role in this endeavour.\\



\acknowledgements

A. J. K. acknowledges research fellowships by the Leopoldina Foundation/Germany (under grant
BMBF-LPD 9901/8-87) and the Swedish Research Council. F. G. acknowledges financial support from the Instrument Center for Danish Astrophysics (IDA) and the Carlsberg Foundation. O. R. thanks the Centre Informatique National de l'Enseignement Sup\'{e}rieur (CINES) and the R\'{e}seau Qu\'{e}b\'{e}cois de Calcul de Haute Performance (RQCHP) for providing the computational resources required for this work.  L. M. acknowledges support through the Russian Academy of Science Presidium Programme `Origin and evolution of stars and the Galaxy'. The Uppsala group of authors acknowledges support from the Swedish Research Council. We thank A.~Alonso and I.~Ram\'{i}rez for providing colour-temperature relations specific to this project.\\



\noindent
{\it Facilities:} \facility{VLT:Kueyen (FLAMES+UVES)}



\clearpage

\appendix

\section{Notes on individual elements}
\subsection{Lithium}
\begin{figure}[!t]\centering
\includegraphics[angle=90,scale=0.4]{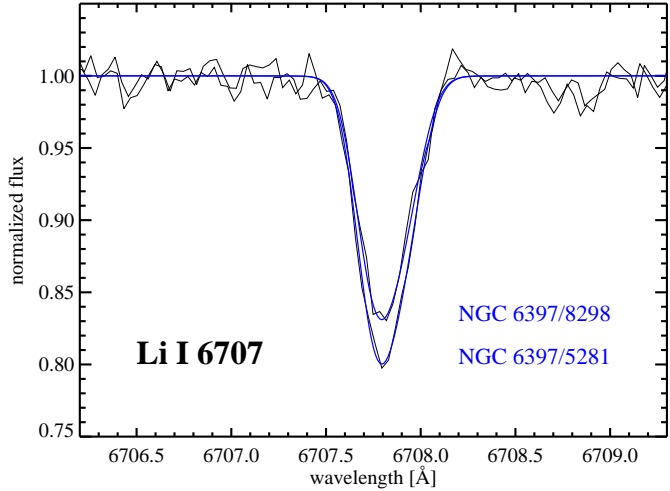}
\caption{Observed Li\,{\sc i} 6707 line profiles for the two SGB stars. These two stars seem to have slightly (0.08\,dex) different lithium abundances. With only two stars in this evolutionary stage it is, however, impossible to tell whether this indicates non-negligible dispersion.}
\label{lithiumSGB}
\end{figure}

We adopt 0.1\,dex as a representative error for the {\em absolute} lithium abundance. However, the star-to-star scatter among the TOP stars is smaller: at 0.05\,dex (1$\sigma$) it can likely be fully ascribed to observational uncertainties. Line strengths vary between 21.3\,m\AA\ (star 12318) and 26.5\,m\AA\ (star 506120).

We use abundances in LTE, as the combined correction from 3D and NLTE was found to be very small \citep{Barklem_etal_2003}. NLTE corrections in 1D are typically $-$0.1\,dex for both TOP and SGB stars when charge-exchange reactions are considered. They decrease to $-$0.07\,dex for the bRGB stars and further to $-$0.04\,dex for the RGB stars. The need for extra mixing discussed in Section \ref{disc} is thus somewhat alleviated when 1D NLTE abundances are considered. Studies in 3D NLTE will shed further light on this question.

\subsection{Magnesium}
\begin{figure}[!t]\centering
\includegraphics[angle=90,scale=0.4]{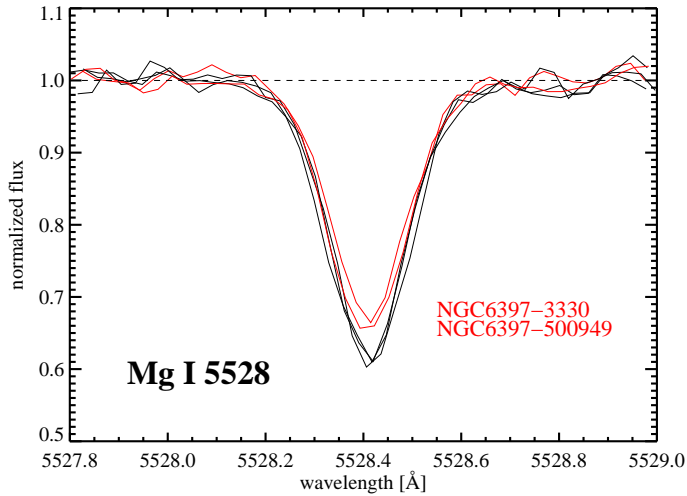}
\caption{Observed Mg\,{\sc i} 5528 line profiles for the five bRGB stars. Both star 3330 and 500949 have significantly weaker profiles than the other three bRGB stars. We attribute this to the conversion of Mg to Al commonly observed in globular clusters (see, e.g., G01).}
\label{magnesium}
\end{figure}

As lines of silicon are not detected in our spectra, we use magnesium as an element for which the stellar-structure models of \citet{Richard_etal_2005} predict trends very similar to silicon ($\Delta \log \varepsilon$\,(Mg) = 0.2, see {\S}\ref{chemabund}). However, magnesium is known to anti-correlate with aluminium in globular clusters. The atomic-diffusion signature could thus be masked by such anti-correlations.

Line strength for Mg\,{\sc i} 5528 vary between 42\,m\AA\ (TOP stars) and 87\,m\AA\ (RGB stars). NLTE effects on Mg\,{\sc i} 5528 vary between +0.11 and +0.14\,dex. Thus, the LTE abundance trend is very similar, albeit at lower overall magnesium abundances.

While we find no obvious line-strength differences for the TOP and SGB stars, both the bRGB and RGB stars show some line-strength variation for Mg\,{\sc i} 5528. Figure \ref{magnesium} clearly shows that two of the five bRGB stars have weaker Mg\,{\sc i} 5528 lines. This is confirmed by NLTE analyses of the Mg\,{\sc i} $b$ lines at 5172 and 5183\,\AA. The overall range of magnesium abundances is found to be 0.2\,dex in both groups of evolved stars. Among the RGB stars, 11093 seems to be the least mixed, displaying the strongest magnesium lines and the weakest line of Na\,{\sc i} 5688.

\subsection{Calcium}
Together with barium, calcium shows NLTE abundance trends that are significantly different from their LTE counterparts. While overionization dominates in the TOP stars (weakening the lines and increasing the abundances by between +0.05 and +0.12\,dex), there are competing NLTE effects between the line core and the line wings as the lines become stronger towards the RGB stars. For example, Ca\,{\sc i} 6162 has an average line strength of 43\,m\AA\ in the TOP stars and the NLTE correction is +0.1\,dex. In the RGB stars, this line is 92\,m\AA\ strong and the NLTE effects is $-$0.05\,dex. Average NLTE effects for all three lines (Ca\,{\sc i} 6122, 6162 and 6439) range from +0.11\,dex (TOP stars) to $-$0.03\,dex (RGB stars). $\Delta \log \varepsilon$\,(Ca)$_{\rm LTE}$ = 0.21, while $\Delta \log \varepsilon\,$(Ca)$_{\rm NLTE}$ = 0.07.

\subsection{Titanium}
The titanium abundance is derived from two weak Ti\,{\sc ii} lines at 5188\,\AA\ and 5226\,\AA, respectively. The prior line is blended with a Vanadium\,{\sc i} line at 5188.85\,\AA, but the resolving power of FLAMES-UVES is high enough to safely separate the contribution of this element. Line strengths vary between 30\,m\AA\ and 80\,m\AA\ (Ti\,{\sc i} 5188) and 22\,m\AA\ and 67\,m\AA\ (Ti\,{\sc i} 5226), respectively. LTE is assumed to be a good approximation for these lines arising from transitions in the dominant ionization stage.

\subsection{Barium}
The analysis of this element is solely based on Ba\,{\sc ii} 6496. Line strengths vary between 8\,m\AA\ (TOP stars) and 66\,m\AA\ (RGB stars). While the LTE trend is steep ($\Delta \log \varepsilon\,$(Ba)$_{\rm LTE}$ = 0.21), no significant trend is found in NLTE ($\Delta \log \varepsilon\,$(Ba)$_{\rm NLTE}$ = 0.00). Like in the case of calcium, NLTE effects change sign in going from the TOP group (+0.11\,dex) to the RGB group ($-$0.10\,dex). The SGB and bRGB groups fall in between in terms of NLTE corrections.

We caution that the TOP-star line is rather weak and may well be affected by telluric absorption.

No atomic-diffusion predictions exist for barium. In the future, efforts should be made to conpare our observations with detailed calculations for the expected behaviour of barium between these groups of stars.

\subsection{Iron}
In Table~7, the lines used for the surface-gravity, iron-abundance and microturbulence determination for the four groups of stars are given. Lines are listed according to their lower excitation energy, the corresponding line data can be found in \citet{Korn_etal_2003}.

\clearpage

\begin{table*}\label{ironlines}
\caption{Iron lines used in the stellar-parameter and iron-abundance determination.}
\begin{center}\small
\begin{tabular}{lcccc}
\tableline\tableline
line & TOP & SGB & bRGB & RGB \\
\tableline
\tableline
Fe\,{\sc i} 5225.5 &            &            & $\bullet$ & $\bullet$ \\
Fe\,{\sc i} 5247.0 &            &            &           & $\bullet$ \\
Fe\,{\sc i} 5250.2 &            &            &           & $\bullet$ \\
Fe\,{\sc i} 5269.5 & $\bullet$ & & & \\
Fe\,{\sc i} 5328.0 & $\bullet$ & & & \\
Fe\,{\sc i} 5371.4 & $\bullet$ & & & \\
Fe\,{\sc i} 5397.1 & $\bullet$  &            & $\bullet$ & \\
Fe\,{\sc i} 5405.7 & $\bullet$  & $\bullet$  & $\bullet$ & \\
Fe\,{\sc i} 5250.6 &            &            &           & $\bullet$ \\
Fe\,{\sc i} 6421.3 &            & $\bullet$  & $\bullet$ & $\bullet$ \\
Fe\,{\sc i} 6663.4 &            &            &           & $\bullet$ \\
Fe\,{\sc i} 6494.9 & $\bullet$ & & & \\
Fe\,{\sc i} 6593.8 &            &            &           & $\bullet$ \\
Fe\,{\sc i} 6065.4 &            & $\bullet$  & $\bullet$ & $\bullet$ \\
Fe\,{\sc i} 6200.3 &            &            &           & $\bullet$ \\
Fe\,{\sc i} 6322.6 &            &            &           & $\bullet$ \\
Fe\,{\sc i} 6546.2 &            &            & $\bullet$ & $\bullet$ \\
Fe\,{\sc i} 6592.9 &            & $\bullet$  & $\bullet$ & $\bullet$ \\
Fe\,{\sc i} 4890.7 & $\bullet$  & $\bullet$  & $\bullet$ & $\bullet$ \\
Fe\,{\sc i} 4891.4 & $\bullet$  & $\bullet$  & $\bullet$ & $\bullet$ \\
Fe\,{\sc i} 4918.9 & $\bullet$  & $\bullet$  & $\bullet$ & $\bullet$ \\
Fe\,{\sc i} 4920.5 & $\bullet$  & $\bullet$  & $\bullet$ & \\
Fe\,{\sc i} 4957.2 & $\bullet$  & $\bullet$  & $\bullet$ & $\bullet$ \\
Fe\,{\sc i} 4957.5 & $\bullet$  & $\bullet$  & $\bullet$ & \\
Fe\,{\sc i} 5139.2 &            &            & $\bullet$ & \\
Fe\,{\sc i} 5139.4 &            &            & $\bullet$ & \\
Fe\,{\sc i} 5232.9 & $\bullet$  & $\bullet$  & $\bullet$ & $\bullet$ \\
Fe\,{\sc i} 5266.5 & $\bullet$  & $\bullet$  & $\bullet$ & $\bullet$ \\
Fe\,{\sc i} 5217.3 &            & $\bullet$  & $\bullet$ & $\bullet$ \\
Fe\,{\sc i} 5253.4 &            &            &           & $\bullet$ \\
Fe\,{\sc i} 5324.1 & $\bullet$  &            & $\bullet$ & $\bullet$ \\
Fe\,{\sc i} 5586.7 & $\bullet$  & $\bullet$  & $\bullet$ & \\
Fe\,{\sc i} 5615.6 &            & $\bullet$  & $\bullet$ & $\bullet$ \\
Fe\,{\sc i} 6232.6 &            &            &           & $\bullet$ \\
Fe\,{\sc i} 6246.3 &            & $\bullet$  & $\bullet$ & $\bullet$ \\
Fe\,{\sc i} 4985.2 &            &            &           & $\bullet$ \\
Fe\,{\sc i} 5074.7 &            &            & $\bullet$ & $\bullet$ \\
Fe\,{\sc i} 5364.8 &            & $\bullet$  & $\bullet$ & $\bullet$ \\
Fe\,{\sc i} 5424.0 & $\bullet$  & $\bullet$  & $\bullet$ & $\bullet$ \\
Fe\,{\sc i} 5410.9 &            & $\bullet$  & $\bullet$ & $\bullet$ \\
Fe\,{\sc i} 6024.0 &            &            &           & $\bullet$ \\
\tableline
Fe\,{\sc ii} 6516.0 &           &           &            & $\bullet$ \\
Fe\,{\sc ii} 5284.1 &           &           &            & $\bullet$ \\
Fe\,{\sc ii} 5169.0 & $\bullet$ & $\bullet$ & $\bullet$ & $\bullet$ \\
Fe\,{\sc ii} 5018.4 & $\bullet$ & $\bullet$ & $\bullet$ & $\bullet$ \\
Fe\,{\sc ii} 4923.9 & $\bullet$ & $\bullet$ & $\bullet$ & $\bullet$ \\
Fe\,{\sc ii} 5362.8 &           &           &           & $\bullet$ \\
Fe\,{\sc ii} 5316.6 & $\bullet$ & $\bullet$ & $\bullet$ & $\bullet$ \\
Fe\,{\sc ii} 5234.6 & $\bullet$ & $\bullet$ & $\bullet$ & $\bullet$ \\
Fe\,{\sc ii} 5197.5 &           & $\bullet$ & $\bullet$ & $\bullet$ \\
Fe\,{\sc ii} 5284.1 &           &           &             & $\bullet$ \\
Fe\,{\sc ii} 6456.3 &           &           &             & $\bullet$ \\
Fe\,{\sc ii} 6247.5 &           &           &             & $\bullet$ \\
\tableline
\end{tabular}
\end{center}
\end{table*}

\clearpage

\section{Photometric calibrations and stellar parameters}

\begin{table*}[!h]
\caption{Photometric calibrations for $v-y$ based on two different calibrations of the infrared-flux method, \citet{Alonso_etal_1996}/\citet{Alonso_etal_1999} and \citet{Ramirez_Melendez_2005}. The coefficients refer to the analytic formulae given in the respective paper.}
\label{photcalib}
\begin{center}\tiny
\begin{tabular}{lrrrrrrrrrrr}
\tableline\tableline
index -- calib.  & a0     & a1    &   a2    &   a3    &   a4   &    a5   &   P0     &      P1     &      P2      &     P3    &    P4 \\
\tableline
  Ram\'{i}rez\\
  (v-y) -- giant & { 0.5067} & { 0.4163} & { -0.0522} &  { 0.0132} & { -0.0984} & { -0.0141} & { 3230.60}   &  { -7214.50}    &  { 5698.73}   &  { -1894.18}  & { 226.107} \\
  (v-y) -- dwarf & { 0.4739} & { 0.4552} & { -0.0556} & { -0.0379} & { -0.0378} & { -0.0186} &  { 269.223}  &   { -514.148}   &   { 234.499}  &      { 0.000} &   { 0.000} \\
  Alonso\\
  (v-y) -- giant & { 0.5408} &  { 0.3074} & { 0.03415} & { -0.00796}  & { -0.05368} & { -0.009701}\\
  (v-y) -- dwarf & { 0.3726} &  { 0.6648} & { -0.1597} &  { -0.03442} & { -0.02909} & { -0.01220}\\
\tableline
\end{tabular}
\end{center}
\end{table*}

\begin{table*}
\caption{Photometric stellar parameters of the 18 stars observed with FLAMES-UVES. The indices are calibrated on the IRFM by Alonso et al.~(1996,1999). At the metallicity of NGC 6397, the calibrations of \citet{Ramirez_Melendez_2005} produce very similar results. For the SGB and bRGB stars, the results of both the dwarf and giant calibration are given.}
\label{table:phot}
\begin{center}\tiny
\begin{tabular}{rccrcrcrcrc}
\tableline\tableline
star id & RA (J2000) & $V$ & $b-y$ & $T_{\rm eff}$/log $g$ & $v-y$ & $T_{\rm eff}$/log $g$ & $B-V$ & $T_{\rm eff}$/log $g$ & $V-I$ & $T_{\rm eff}$/log $g$\\
 & DEC (J2000) & & & dwarf -- giant & & dwarf -- giant & & dwarf -- giant & & dwarf -- giant\\
\tableline
NGC6397-11093   &    17 40 13.950  & 13.551 &   0.478 & \hspace{6.5ex} -- 5102 & 1.052 & \hspace{6.5ex} -- 5104 & 0.651 & \hspace{6.5ex} -- 5224  &  0.875 & \hspace{6.5ex} -- 5034 \\
                &    $-$53 42 25.40  &        &         & \hspace{6ex} -- 2.51 &       & \hspace{6ex} -- 2.51 &       & \hspace{6ex} -- 2.56  &        & \hspace{6ex} -- 2.48 \\
NGC6397-13092   &    17 40 19.980  & 13.644 &   0.478 & \hspace{6.5ex} -- 5103 & 1.047 & \hspace{6.5ex} -- 5115 & 0.646 & \hspace{6.5ex} -- 5242  &  0.868 & \hspace{6.5ex} -- 5051 \\
                &    $-$53 39 17.70  &        &         & \hspace{6ex} -- 2.54 &       & \hspace{6ex} -- 2.55 &       & \hspace{6ex} -- 2.60  &        & \hspace{6ex} -- 2.52 \\
NGC6397-14592   &    17 40 23.840  & 13.696 &   0.476 & \hspace{6.5ex} -- 5113 & 1.041 & \hspace{6.5ex} -- 5126 & 0.642 & \hspace{6.5ex} -- 5251  &  0.865 & \hspace{6.5ex} -- 5060 \\
                &    $-$53 36 40.40  &        &         & \hspace{6ex} -- 2.57 &       & \hspace{6ex} -- 2.58 &       & \hspace{6ex} -- 2.63  &        & \hspace{6ex} -- 2.55 \\
NGC6397-7189    &    17 39 59.760  & 13.729 &   0.474 & \hspace{6.5ex} -- 5120 & 1.038 & \hspace{6.5ex} -- 5134 & 0.641 & \hspace{6.5ex} -- 5257  &  0.862 & \hspace{6.5ex} -- 5066 \\
                &    $-$53 45 03.90  &        &         & \hspace{6ex} -- 2.59 &       & \hspace{6ex} -- 2.59 &       & \hspace{6ex} -- 2.64  &        & \hspace{6ex} -- 2.56 \\
NGC6397-4859    &    17 39 48.710  & 13.815 &   0.470 & \hspace{6.5ex} -- 5139 & 1.028 & \hspace{6.5ex} -- 5153 & 0.636 & \hspace{6.5ex} -- 5272  &  0.857 & \hspace{6.5ex} -- 5081 \\
                &    $-$53 40 16.90  &        &         & \hspace{6ex} -- 2.63 &       & \hspace{6ex} -- 2.63 &       & \hspace{6ex} -- 2.68  &        & \hspace{6ex} -- 2.60 \\
NGC6397-502074  &    17 40 50.850  & 13.853 &   0.468 & \hspace{6.5ex} -- 5148 & 1.024 & \hspace{6.5ex} -- 5162 & 0.634 & \hspace{6.5ex} -- 5279  &  0.854 & \hspace{6.5ex} -- 5088 \\
                &    $-$53 36 08.80  &        &         & \hspace{6ex} -- 2.65 &       & \hspace{6ex} -- 2.65 &       & \hspace{6ex} -- 2.70  &        & \hspace{6ex} -- 2.62 \\
NGC6397-3330    &    17 39 40.250  & 15.227 &   0.428 & 5405 -- 5351 & 0.932 & 5436 -- 5359 & 0.568 & 5492 -- 5491  &  0.784 & 5368 -- 5281 \\
                &    $-$53 41 49.60  &        &         & 3.30 -- 3.28 &       & 3.31 -- 3.28 &       & 3.33 -- 3.33  &        & 3.28 -- 3.25 \\
NGC6397-23267   &    17 40 43.750  & 15.339 &   0.424 & 5429 -- 5373 & 0.920 & 5466 -- 5385 & 0.560 & 5520 -- 5519  &  0.778 & 5386 -- 5298 \\
                &    $-$53 37 17.40  &        &         & 3.35 -- 3.33 &       & 3.37 -- 3.34 &       & 3.39 -- 3.39  &        & 3.34 -- 3.30 \\
NGC6397-15105   &    17 40 25.540  & 15.439 &   0.419 & 5458 -- 5399 & 0.909 & 5497 -- 5411 & 0.551 & 5552 -- 5551  &  0.770 & 5410 -- 5321 \\
                &    $-$53 43 31.10  &        &         & 3.40 -- 3.38 &       & 3.42 -- 3.39 &       & 3.44 -- 3.44  &        & 3.39 -- 3.35 \\
NGC6397-500949  &    17 40 34.580  & 15.514 &   0.416 & 5480 -- 5418 & 0.900 & 5520 -- 5431 & 0.544 & 5576 -- 5574  &  0.764 & 5428 -- 5338 \\
                &    $-$53 33 20.70  &        &         & 3.44 -- 3.42 &       & 3.46 -- 3.42 &       & 3.48 -- 3.48  &        & 3.42 -- 3.39 \\
NGC6397-6391    &    17 39 55.900  & 15.551 &   0.412 & 5503 -- 5437 & 0.890 & 5547 -- 5452 & 0.539 & 5593 -- 5591  &  0.760 & 5442 -- 5351 \\
                &    $-$53 35 11.90  &        &         & 3.47 -- 3.44 &       & 3.48 -- 3.45 &       & 3.50 -- 3.50  &        & 3.44 -- 3.41 \\
NGC6397-8298    &    17 40 04.130  & 15.832 &   0.369 & 5791 -- 5677 & 0.799 & 5819 -- 5666 & 0.478 & 5822 -- 5819  &  0.689 & 5682 -- 5571 \\
                &    $-$53 40 53.10  &        &         & 3.68 -- 3.64 &       & 3.69 -- 3.63 &       & 3.69 -- 3.69  &        & 3.64 -- 3.60 \\
NGC6397-5281    &    17 39 50.570  & 15.839 &   0.368 & 5802 -- 5687 & 0.796 & 5828 -- 5673 & 0.476 & 5832 -- 5828  &  0.686 & 5693 -- 5580 \\
                &    $-$53 35 21.70  &        &         & 3.68 -- 3.64 &       & 3.69 -- 3.64 &       & 3.69 -- 3.69  &        & 3.65 -- 3.61 \\
NGC6397-10197   &    17 40 10.770  & 16.160 &   0.314 & 6204 -- \hspace{6.5ex} & 0.695 & 6187 -- \hspace{6.5ex} & 0.402 & 6135 -- \hspace{6.5ex}  &  0.585 & 6096 -- \hspace{6.5ex} \\
                &    $-$53 38 26.40  &        &         & 3.93 -- \hspace{6ex} &       & 3.93 -- \hspace{6ex} &       & 3.91 -- \hspace{6ex}  &        & 3.90 -- \hspace{6ex} \\
NGC6397-12318   &    17 40 17.640  & 16.182 &   0.312 & 6217 -- \hspace{6.5ex} & 0.691 & 6201 -- \hspace{6.5ex} & 0.399 & 6148 -- \hspace{6.5ex}  &  0.581 & 6113 -- \hspace{6.5ex} \\
                &    $-$53 39 34.20  &        &         & 3.95 -- \hspace{6ex} &       & 3.94 -- \hspace{6ex} &       & 3.93 -- \hspace{6ex}  &        & 3.92 -- \hspace{6ex} \\
NGC6397-9655    &    17 40 09.090  & 16.200 &   0.311 & 6227 -- \hspace{6.5ex} & 0.688 & 6212 -- \hspace{6.5ex} & 0.396 & 6159 -- \hspace{6.5ex}  &  0.578 & 6128 -- \hspace{6.5ex} \\
                &    $-$53 43 26.40  &        &         & 3.96 -- \hspace{6ex} &       & 3.95 -- \hspace{6ex} &       & 3.94 -- \hspace{6ex}  &        & 3.93 -- \hspace{6ex} \\
NGC6397-506120  &    17 40 41.590  & 16.271 &   0.308 & 6248 -- \hspace{6.5ex} & 0.683 & 6233 -- \hspace{6.5ex} & 0.391 & 6179 -- \hspace{6.5ex}  &  0.570 & 6163 -- \hspace{6.5ex} \\
                &    $-$53 45 49.30  &        &         & 3.99 -- \hspace{6ex} &       & 3.99 -- \hspace{6ex} &       & 3.97 -- \hspace{6ex}  &        & 3.97 -- \hspace{6ex} \\
NGC6397-507433  &    17 40 16.070  & 16.278 &   0.308 & 6249 -- \hspace{6.5ex} & 0.682 & 6235 -- \hspace{6.5ex} & 0.391 & 6180 -- \hspace{6.5ex}  &  0.570 & 6166 -- \hspace{6.5ex} \\
                &    $-$53 47 18.60  &        &         & 3.99 -- \hspace{6ex} &       & 3.99 -- \hspace{6ex} &       & 3.97 -- \hspace{6ex}  &        & 3.97 -- \hspace{6ex} \\
\tableline
\multicolumn{3}{l}{$\Delta T_{\rm eff} ({\rm TOP} - {\rm RGB})$} & & 1108 & & 1082 & & 906 & & 1070 \\
\multicolumn{3}{l}{$\Delta \log g ({\rm TOP} - {\rm RGB})$} & & 1.38 & & 1.37 & & 1.31 & & 1.38 \\
\tableline
\end{tabular}
\end{center}
\end{table*}

\clearpage





\end{document}